\documentclass[twocolumn,twocolappendix]{aastex63}

\newcommand\psscore{{\it ps\/}-score}
\newcommand\psscores{{\it ps\/}-scores}

\usepackage{bm}
\usepackage{todonotes}

\received{August 17, 2020}
\revised{October 5, 2020}
\accepted{October 16, 2020}

\submitjournal{AJ}

\shorttitle{Improvements to Pan-STARRS1 Astrometry }
\shortauthors{Lubow, White, \& Shiao}
\begin{document}

\title{Improvements to Pan-STARRS1 Astrometry Using Gaia}

\correspondingauthor{Stephen Lubow}
\email{lubow@stsci.edu}

\author[0000-0002-4636-7348]{Stephen H. Lubow}
\author[0000-0002-9194-2807]{Richard L. White}
\author[0000-0001-7842-3714]{Bernie Shiao}
\affiliation{Space Telescope Science Institute, 3700 San Martin Drive, Baltimore, MD 21218, USA}

\begin{abstract}
	We use the Gaia DR2 catalog to improve the astrometric accuracy of about 1.7 billion objects in Pan-STARRS1 Data Release 2 (PS1 DR2). We also obtain proper motions for these PS1 objects. The cross match between Gaia and PS1 reveals residuals that are correlated
on a scale of about 1 arcmin. We apply a spatially adaptive correction algorithm for all PS1 objects having more than two detections to reduce these residuals and align the object positions to Gaia.
For point-like PS1 objects
that cross match to Gaia, the algorithm reduces PS1/Gaia residuals by 33\% in position (median value of 13.5 mas reduced to 9.0 mas) and by 24\% in proper motion (median value of 6.3 mas/yr reduced to 4.8 mas/yr). The residuals for the corrected positions are smallest for objects with the most point-like morphologies and with intermediate magnitudes of about 17 mag.  
The residual errors in declination are systematically larger than those in right
ascension; the declination errors increase with zenith angle in proportion to
the air mass of the observations.  Declination positional residuals at a given
declination generally vary with color and are consistent with the effects of
differential atmospheric refraction.  In principle, these residuals could be
reduced  further by taking into account object color.

\end{abstract}


\section{Introduction}

The Panoramic Survey Telescope and Rapid Response System (Pan-STARRS) has carried out a 3$\pi$ steradian survey of the sky, north of declination ${-}30$ degrees.
The Pan-STARRS1 survey (PS1) was carried out mainly from 2010 to 2014 at Haleakala Observatory in Hawaii by covering the sky region about 12 times
in each of five broadband filters  ($g$, $r$, $i$, $z$, $y$).
Several papers are available that describe the system design, pipeline, calibration, and results; \cite{Chambers2016} provides an overview.
There have been two data releases, DR1 and DR2.
DR1 (2016 December) contained only average information resulting from individual exposures. The second data release, DR2 (2019 January), contains time-dependent information obtained from individual exposures.
In this paper we make use of DR2.\footnote{When we refer to PS1 data products, we mean the DR2 versions unless we explicitly mention
DR1.}

PS1 DR2 contains over 1.6 petabytes of data including measurements for more than 10 billion objects, making it the largest volume of astronomical information released to date. The image data and an object catalog are made available online at \url{http://panstarrs.stsci.edu} through the \textit{Barbara A. Mikulski Archive for Space Telescopes (MAST)} at the Space Telescope Science Institute.
The catalog consists of tables that are stored in Microsoft SQL Server databases \citep{Flewelling2016}.
The PS1 catalog database makes use of software previously developed for SDSS on SQL Server \citep{Thakar2003, Heasley2008}. The tables can be accessed through a web form and Application Programming Interface (API) at \url{https://catalogs.mast.stsci.edu} or through
an SQL query interface called CasJobs at \url{https://mastweb.stsci.edu/ps1casjobs}.  It is also accessible via a Virtual Observatory Table Access Protocol\footnote{\url{http://ivoa.net/documents/TAP}} (TAP) interface at \url{http://vao.stsci.edu/PS1DR2/tapservice.aspx}.

In this paper we are concerned with improving
the astrometry of the PS1 DR2 catalog.
There are several motivations to improve the
PS1 astrometry.
Although the Gaia catalog \citep{Lindegren2018} provides more accurate astrometry than PS1, PS1 reaches fainter objects (mag 23) than Gaia (mag 20). For this reason,
PS1 astrometry can play an important role.
For example, for the purposes of determining the astrometry of images taken by the \textit{Hubble Space Telescope},
we have found in the construction of the Hubble Source Catalog \citep{Whitmore2016} that about 20\% to 30\% of the images
taken with the wide field instruments
(the Wide Field and Planetary Camera 2, Advanced Camera for Surveys, and Wide Field Camera 3)
do not contain enough unsaturated stars in common with
Gaia to provide an astrometric solution.
This situation is of importance for high Galactic latitude fields where the density of stars is relatively low.
The \textit{James Webb Space Telescope} (JWST) will have similar issues, with the added complication that even faint Gaia stars will
often be saturated in JWST images.
In such cases, PS1 provides an important alternative astrometric reference catalog having higher source density and fainter objects.

The astrometric calibration of the current PS1 DR2 catalog is described in \cite{Magnier2016}.
The astrometry was calibrated using 2MASS \citep{Skrutskie2006} and Gaia DR1 \citep{Lindegren2016}
object positions that were available at the time of the PS1 data processing. The Gaia positions were much more heavily weighted in the PS1 astrometric determinations due to their much smaller errors.  A major issue in this calibration procedure is the presence of systematic proper motions that lead to nonrandom shifts between PS1 stellar positions and those in 2MASS and Gaia. Gaia DR1 does not provide proper motions.
These proper motions were modeled in PS1  by using by a flat rotation curve for the Galactic disk. Making these proper motion corrections for each cross-matched disk object requires a knowledge of the distance to the object that is estimated by its distance modulus. As discussed by \cite{Magnier2016}, this process and other issues introduce some astrometric  uncertainties. In this paper we utilize the Gaia DR2 positions, proper motions, and parallaxes to improve the PS1 astrometry.

In comparing Gaia and PS1 object positions, we find small scale $\sim 1$ arcmin structures in the residuals. To mitigate against such effects, we employ an adaptive astrometric correction algorithm that smooths and reduces residuals on that scale.

In this paper we describe the determination for the first time of  proper motions for PS1 objects.
We describe the correction algorithm for improving the astrometric accuracy of PS1
by making use of Gaia DR2 and analyze the results.
We also apply a similar correction procedure to improve the PS1 proper motion
accuracy by making use of the proper motions available
in Gaia DR2. These improvements will be integrated into the
MAST PS1 catalog.

The outline of this paper is as follows.
In Section \ref{sec:background} we provide some background for our calculations by describing the mean object positions and proper motions that we utilize as inputs to our correction algorithm. We also provide
some examples of the PS1/Gaia residuals that we aim to correct. Section \ref{sec:algorithm} describes our correction algorithm.
Section \ref{sec:global} describes the PS1 positional
and proper motion residuals of reference objects (high quality detections that cross match to Gaia) that occur over the entire PS1 sky region before and after the corrections we apply.
In Section \ref{sec:objprop} we describe how the number of the objects varies with declination.
We then discuss how the position residuals of reference objects
vary with declination (Section~\ref{sec:posresw}) and color (Section \ref{sec:color}).
Section \ref{sec:pmresidstr} describes the variations
of proper motion residuals
with declination. Section \ref{sec:checks}
describes some additional checks on the results, including comparisons to
the Hubble Source Catalog and the
ICRF2 radio catalog \citep{Ma2009} to explore the PS1 astrometric accuracy at
faint magnitudes.
Section \ref{sec:summary} contains the discussion and summary.
The convergence of the positional correction shifts
is discussed in Appendix \ref{sec:convergence}.

\section{Background}
\label{sec:background}

The PS1 database tables contain
 positional and photometric information
about the detected objects. The database contents are described in detail in
\cite{Flewelling2016} and in the PS1 archive documentation (\url{https://panstarrs.stsci.edu}).
Positional information based on individual single-epoch exposures is available in the
\textit{Detection} table (actually, a view or virtual table).
Stack images are creating by combining all the single-epoch exposures in a given filter to get a
deeper image.
Table \textit{StackObjectThin} contains positional
information on each object for each filter as determined from the stack images.
The \textit{ObjectThin} table contains summary information from both the single-epoch measurements and the stack measurements
in the form of weighted means of the values measured at different epochs and filters.

\subsection{Computation of Mean Positions}
\label{sec:computation-of-mean}

It might appear natural to recalibrate the astrometry starting from the weighted positions (raMean and decMean) in the \textit{ObjectThin} table.  However, that is not
possible because those positions are contaminated by Gaia DR1 data for objects that match Gaia.
In PS1 DR1 and DR2, cross-matched Gaia positions are included with high weighting in the determination of mean PS1 positions \citep{Magnier2016}.  The combined positions are weighted by the inverse variances of the position measurements; since the Gaia errors are much smaller than the PS1 errors, the effect is that the database positions for PS1 objects that match Gaia are almost equal to the Gaia DR1 measurements.
A comparison of Gaia and PS1 DR2 positions for the objects in common yields a typical residual of about 5 mas.  While those positions
are in fact very accurate, they derive that accuracy not from the PS1 measurements but from Gaia.  The excellent agreement between PS1 database positions and Gaia positions is \textit{not} an independent measurement of the PS1 positional accuracy, and the PS1/Gaia differences do \textit{not} reflect the typical PS1 astrometric errors for the large majority of PS1 objects that do not match Gaia.
\cite{Makarov2017} carried out
an independent assessment of PS1 astrometry using 19 million Gaia stars that were excluded from the PS1 astrometric calibration
due to their Gaia catalog flags.
They found that the residual systematic uncertainties for the PS1 positions are closer to 20 mas. 
Note that the \cite{Makarov2017} paper relied
on an earlier astrometric calibration of PS1 (PV3.2) that was affected by a poor quality astrometric flag-field correction.
That correction was repaired in PS1 DR2 \citep{Magnier2016}.

Rather that use the positions from \textit{ObjectThin}, we compute new mean positions and proper motions
directly from the single-epoch PS1 positions
in the \textit{Detection} table.
The values in the \textit{Detection} table do not include Gaia DR1 data (although they were calibrated using that data).
We restrict our sample to objects that have good measurements for at least three epochs.  We exclude measurements
having \texttt{infoFlag3} bit 16 set (indicating the measurement was an astrometric outlier) and also exclude detections
having a \texttt{psfQfPerfect} value less than 0.9 (indicating that there are bad pixels within the core of the stellar image).
The full PS1 \textit{ObjectThin} contains entries for about 10 billion objects. After excluding objects with fewer than three good
measurements, we are left with approximately 1.7 billion objects.  Those objects are the topic of this paper.

Note that the astrometric correction algorithm described in this paper also relies on the PS1/Gaia cross-match.
However, for each PS1 object, we explicitly exclude the Gaia match for that object (if there is one); we compute
the correction using only the offsets from nearby objects.  As a result, the positions in our catalog are in the
Gaia reference frame but are not biased by the inclusion of Gaia data.  Our new PS1 positions are independent
measurements that are not correlated with Gaia DR2 measurements of the same stars.

\subsection{Selection of Point-like Reference Objects Using \psscore}

Using a machine learning model, \cite{Tachibana2018} classified about 1.5 billion objects from the PS1 DR1 as either extended objects (resolved) or point sources (unresolved).  Objects were ranked
probabilistically by a quantity called the point-source score (hereafter \psscore) on a scale in which 0 represents extended objects and 1 represents point sources.
The results of this analysis are provided as a MAST high level product (see \url{https://archive.stsci.edu/prepds/ps1-psc/})
and as a CasJobs table in context HLSP\_PS1\_PSC.

In this paper, we make use of \psscore\ values in several ways. We restrict the sample of PS1/Gaia matches used for
the calibration to PS1 objects having high \psscore\ values, which improves the accuracy of the results (particularly in
crowded fields where many PS1 objects are blended).  We also use the \psscore\ values to select sources for assessing the accuracy
of our corrected astrometry.  Our justifications for this approach are discussed below (Section~\ref{sec:global}).

However, we are using PS1 DR2 rather than DR1 for which \psscore\ values were obtained.
PS1 identifies objects with a unique identifier
called \texttt{objid}. The \psscores\ are associated with an \texttt{objid} in DR1. Most of the objects in PS1 DR1 are carried
over with the same \texttt{objid} in DR2, although some DR2 objects do not have a \psscore\ value.
We have found that less than about 10\% of the DR2 objects
are missing \psscores\ in low density regions. The percentage is higher in high density regions such as the Galactic plane. But since there are many PS1 and Gaia sources available
in such regions, our analysis is not much affected
by these missing scores.
Consequently, we are able to make
use of the PS1 DR1 \psscore\ values for certain aspects of our calculations.

Note that we calculate improved astrometry for all 1.7 billion objects selected
as described in Section~\ref{sec:computation-of-mean}, regardless of their
\psscore\ values.  The restriction of our sample using the \psscore\ values
applies only to selecting objects for the positional calibration and for assessing of
the results.

\begin{figure*}
	\centering \leavevmode
    \includegraphics[width=0.49\linewidth]{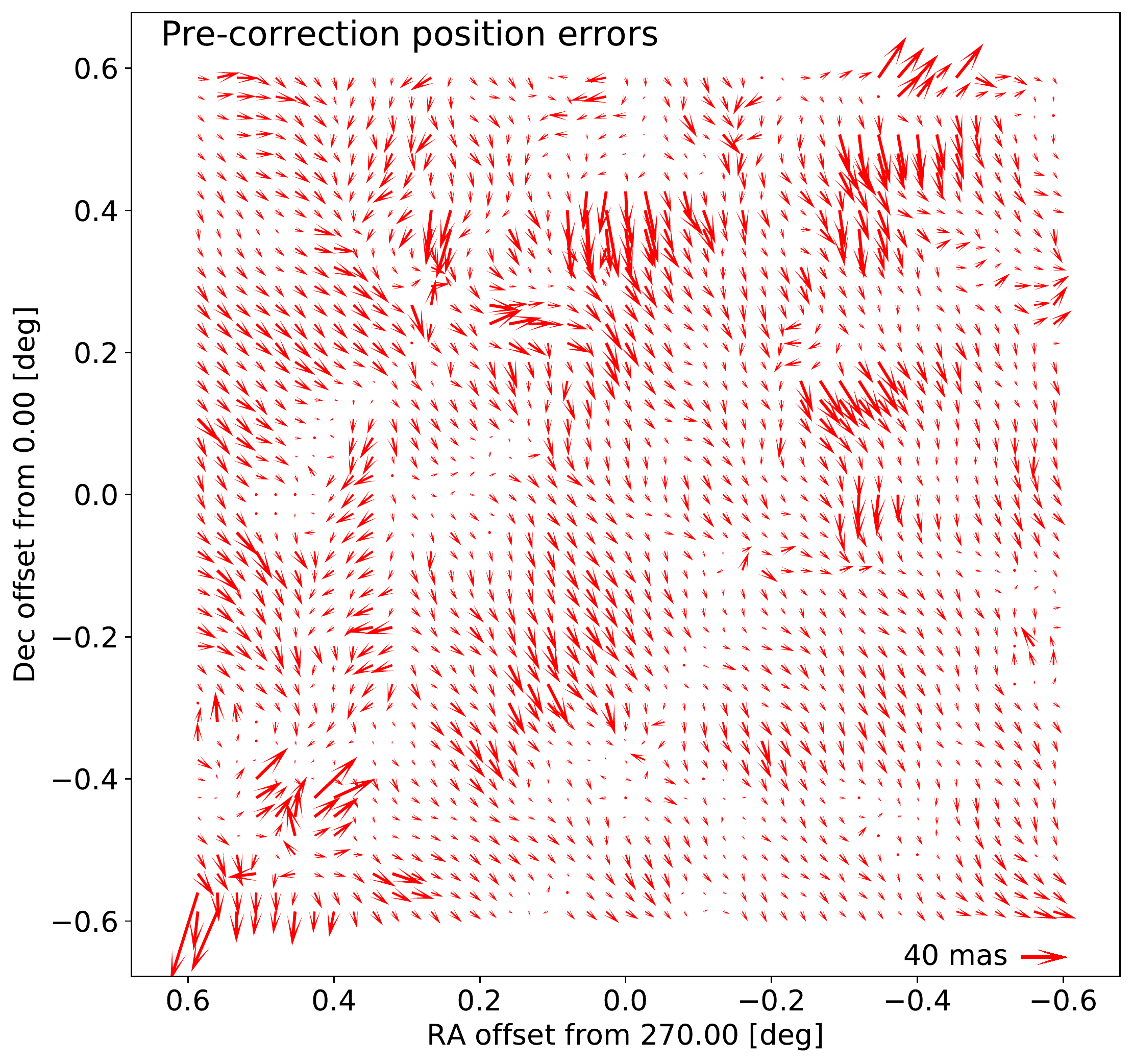}\hfil
    \includegraphics[width=0.49\linewidth]{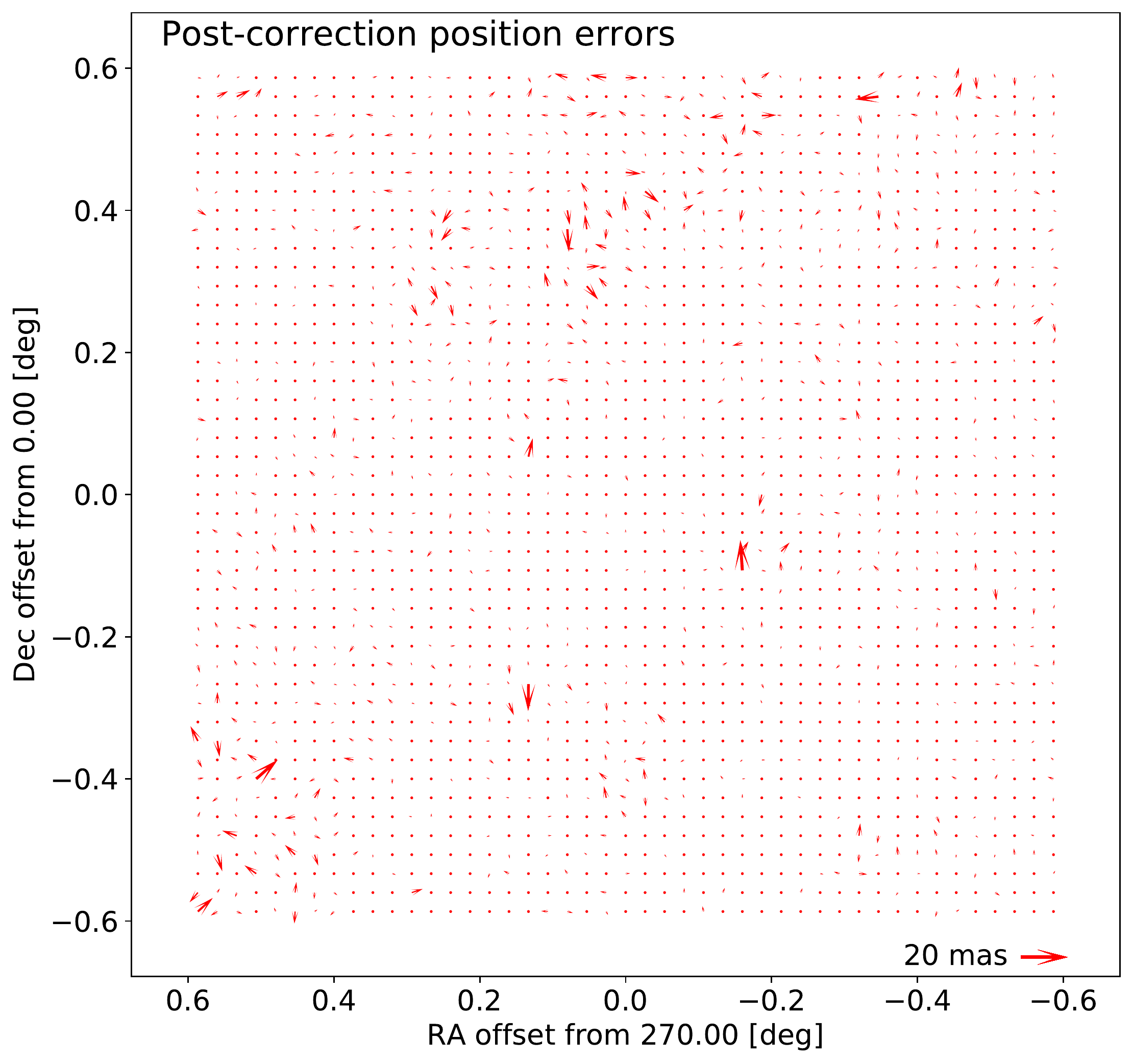}
\caption{Systematic astrometric distortion as a function of position for a
    typical PS1 field. For a moderately crowded field centered at RA
    $270^\circ$, declination $0^\circ$ ($l=27^\circ$, $b=11^\circ$), 92,583 PS1/Gaia
    matches with PS1 point-source scores $>0.9$ \citep{Tachibana2018} are binned
    into cells with $\sim50$ matches each.  The vectors show the median shifts
    between the PS1 and Gaia position in each cell; the scale for the vectors is
    shown at the lower right.  Before the corrections derived in this paper
    (left panel), there are systematic, spatially variable distortions in the
    PS1 coordinates, with a mean shift of 12.7 mas.  After correction (right
	panel) the remaining systematic distortions are small ($<1\,\hbox{mas}$).
	The vector scale has been increased by a factor of two in the right
	panel to increase the visibility of the residuals.}
\label{fig:distortion-example}
\end{figure*}

\begin{figure}
	\centering
\plotone{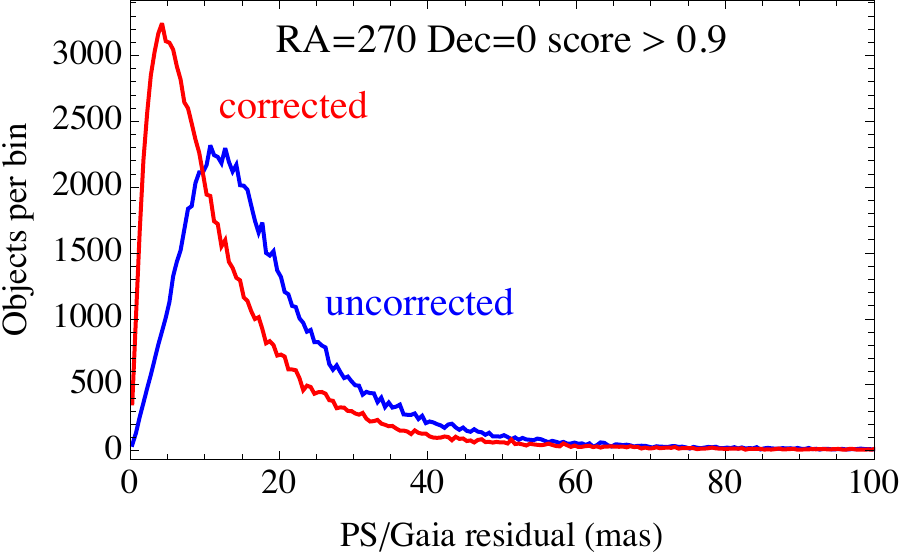}
	\caption{Histograms of the PS1/Gaia position residuals before and after correction for the
	PS1/Gaia sample from Fig.~\ref{fig:distortion-example}.  The removal of systematic
	errors by the procedure described in this paper leads to much improved PS1 astrometry. The bin size is 0.5 mas.}
	\label{fig:distortion-example-hist}
\end{figure}

\subsection{Astrometric Errors in the PS1 DR2 Catalog}

Figures \ref{fig:distortion-example} and \ref{fig:distortion-example-hist}
illustrate the systematic astrometric errors that are corrected in this paper.  The
left panel in Figure~\ref{fig:distortion-example} shows the offsets between our
PS1 mean positions and Gaia DR2 positions as a function of position in a typical
PS1 field.  The Gaia
positions have been shifted using the Gaia proper motions to the epoch of the
corresponding
PS1 object. (The epoch is different for every object due to the complex grid of
overlapping exposures.)  Systematic distortions of up to tens of milliarcseconds are
apparent on scales of one to a few arcminutes.  After correction using the
algorithm described below, the systematic errors are greatly reduced (right
panel).

We are not certain what causes the astrometric
errors on the $\sim 1$ arcmin scale that we correct.
The astrometric calibration carried out by \cite{Magnier2016} involves
making global corrections involving at most cubic polynomials on the chip scale.
Since the chip size is 20 arcmin on each edge, this provides independent corrections on
the larger $\sim 8$ arcmin scale. That is a likely contributor to the corrected
errors.

Figure \ref{fig:distortion-example-hist} shows the distribution of positional
errors before and after our correction for all the objects in this field.  This
field was selected as a typical example of the improvement found
throughout the PS1 survey area.  The pattern and amplitude of the distortions
varies from field to field, but qualitatively similar distortions are found in
all fields.
Removing the systematic, spatially correlated
distortions results in substantial improvements to the overall accuracy of the
PS1 astrometry.  The algorithm used to achieve this improvement is described in the
next section.

\section{Correction Algorithm}
\label{sec:algorithm}

\subsection{Mean Detection Positions}
\label{sec:meandetection}

The initial step in calculating new positions is to combine the positions
measured at every epoch (from the \textit{Detection} table) to get a new ``mean
detection'' position for every object.  As was mentioned above, the
\texttt{infoFlag3} and \texttt{psfQfPerfect} fields are used to exclude poor
quality measurements, and only objects with at least three good measurements are
included.  The \textit{Detection} table's \texttt{ra} and \texttt{dec} values
are combined using $1/\sigma^2$ weighting from the errors \texttt{raErr} and
\texttt{decErr} to get the weighted mean position.  The PMs in RA and
declination are also determined using a weighted linear fit to the positions as
a function of the observing epoch \texttt{obsTime}.  The result of the fit is
a new table of PS1 DR2 positions uncontaminated by the Gaia DR1 measurements
(\texttt{mdra}, \texttt{mddec}), along with the PMs (\texttt{mdpmra},
\texttt{mdpmdec}).  Simple propagation of errors is used to calculate the errors
on these quantitites (\texttt{mdraErr}, \texttt{mddecErr}, \texttt{mdpmraErr},
\texttt{mdpmdecErr}).  The weighted mean observation epochs (\texttt{mdmjdra},
\texttt{mdmjddec}) are used as the reference time for the fits so that the
covariances between the positions and PMs are formally zero.  The $\chi^2$
values of the fits to the data are also computed (\texttt{chisqra}, \texttt{chisqdec}).

\subsection{Astrometric Distortion Correction}
\label{sec:astrometricdistortion}

We correct the systematic distortions between PS1 coordinates and Gaia
coordinates (Fig.~\ref{fig:distortion-example}) using the shifts of nearby PS1
objects that cross match to Gaia, objects that we call \textit{reference
objects}. In doing so, we are assuming that these reference shifts vary
smoothly with position on the scale of the distance between the reference
objects.  This local shift can be computed regardless of whether a particular
PS1 object has a Gaia counterpart or not. (Most PS1 objects do not have a
match in Gaia.)

We experimented with several different algorithms for choosing the cohort of nearby
reference objects (referred to as the \textit{neighborhood}) for correcting the astrometry.
An algorithm using objects out to some fixed radius works well as long as there
are enough close object, but it can fail either when the radius is too small
(leading to too few reference objects and a noisy correction) or when the radius is
too large (leading to a less than optimal correction and a more costly computation
in densely populated regions).  An algorithm that adjusts the radius based on the local
source density is effective but is more complex to implement.

We adopt an algorithm where the neighborhood includes the nearest $N=33$
reference objects, excluding the object itself.  By choosing a fixed number of
neighboring reference objects rather than using all neighbors within a fixed
radius, the algorithm is spatially adapted to the local sky density of
objects.  That produces better results in high density regions while avoiding
noisier corrections from using too few reference objects in low density
regions.  The value of 33 for the number of nearest reference objects was
determined empirically by comparing results for a wide range of values in a
variety of fields that have very different source densities. This number of
nearest neighbors provides accurate Gaia corrections while removing the
small scale structures in the residuals, as we show in Section
\ref{sec:global}.  In Appendix \ref{sec:convergence} we show that this number of
neighbors provides well converged results.

\subsection{Correction Algorithm Steps}
\label{sec:algorithmsteps}

The PS1 coordinate corrections are determined by the steps listed below.

1. For each Gaia DR2 source, the nearest PS1 object with more than two detections is determined within a 2 arcsecond search radius without accounting for corrections due to proper motions and parallaxes.
We utilize the \texttt{raMean} and \texttt{decMean} positions from the \textit{ObjectThin} table for this cross-match.
Those positions include contributions from Gaia DR1 measurements (Section \ref{sec:computation-of-mean}), which for this
cross-matching purpose is not a problem.
In the rare cases
where a PS1 object matches more than one Gaia source, the nearest Gaia source is regarded as the cross-matched source. There
is then a unique Gaia source for each cross-matched PS1 object.

2. We retain high-quality, point-like PS1 objects that have declination $\delta>{-}30$ degrees and a \psscore\ $>0.9$.
Note that all of these PS1 objects have a coordinate and proper motion determination (computed as described
in Section~\ref{sec:meandetection}).
Also, we only include PS1 objects for which the matching Gaia object has a proper motion value.

3. For each reference PS1 object, we compute the Gaia RA and declination at the PS1 RA and declination
mean detection epochs respectively (\texttt{mdmjdra}, \texttt{mdmjddec}) of the cross-matched Gaia source by using the Gaia proper motions and parallaxes.

4. For each reference PS1 object, we compute the RA and declination offsets from the PS1 position to the Gaia position obtained
in Step 3. This set of shift values then provides the reference shifts for making the PS1 astrometry corrections.

5. For each PS1 object (including both reference and non-reference objects), the correction shifts are obtained as the median of the RA and declination reference shifts determined in Step 4 for the nearest 33 reference objects.
If the object being corrected is itself a reference object, it is \textit{excluded} from the sample of reference objects.

6. For each PS1 object, the RA and declination shifts determined in Step 5 are added to the PS1 RA and declination values respectively to obtain the
Gaia-corrected coordinate values.

We apply a similar procedure to correct the PS1 proper motions. The proper motion correction for a PS1 object involves
determining the median proper motion offset (Gaia PM motion minus PS1 PM) of the nearest 33 PS1 reference objects, excluding
the object itself. That PM offset is added to each PS1 object PM to determine the Gaia-corrected proper motion.

These steps were carried out in a Microsoft SQL Server database using the JHU spherical library \citep{Budavari2010} for finding nearest neighbors and using  Common Language Runtime (CLR) functions
for computing the median values and Gaia parallax shifts. The running time for all PS1 objects we consider on Steps 2 to 6 was less than three days.

In total, there are about 1.7 billion objects that have corrected positions and proper motions
based on about 428 million reference objects.

\section{Global Results}
\label{sec:global}

We describe the results of applying the correction algorithm over the entire region of the sky covered by PS1. Recall that these corrections  are based on median  PS1 to Gaia  shifts of objects near each PS1 object we consider, but do not involve the shift of the object being corrected. These residuals then
do not involve a direct fit of each PS1 object to its cross-matched Gaia object.  The residuals provide a measure of the true PS1 error relative to Gaia.
We test this point in Section \ref{sec:rescont}.

The reference objects
used in the algorithm of Section \ref{sec:algorithm} have more than two detections, a \psscore\ that is greater than 0.9, and cross match with Gaia. They have significantly improved astrometry by the corrections we apply.
About 64\% of PS1 objects that have more than two detections and cross match with Gaia are reference objects.
About 38\% of PS1
objects with more than two detections have \psscores\ in this range.
For this section we focus
on the properties of the reference objects.

\begin{figure}
\centering
\includegraphics[width=0.95\columnwidth]{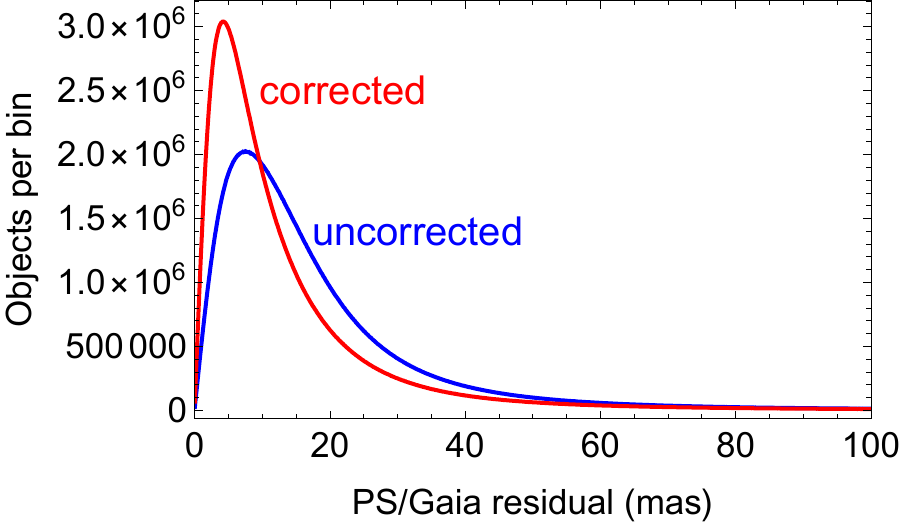}
\caption{Distribution of PS1/Gaia positional residuals in mas with and without the PS1 corrections described in Section \ref{sec:algorithm} for PS1 reference objects. The bin size is 0.1 mas.}
\label{fig:globald}
 \end{figure}

\begin{figure}
\centering
\includegraphics[width=0.907\columnwidth]{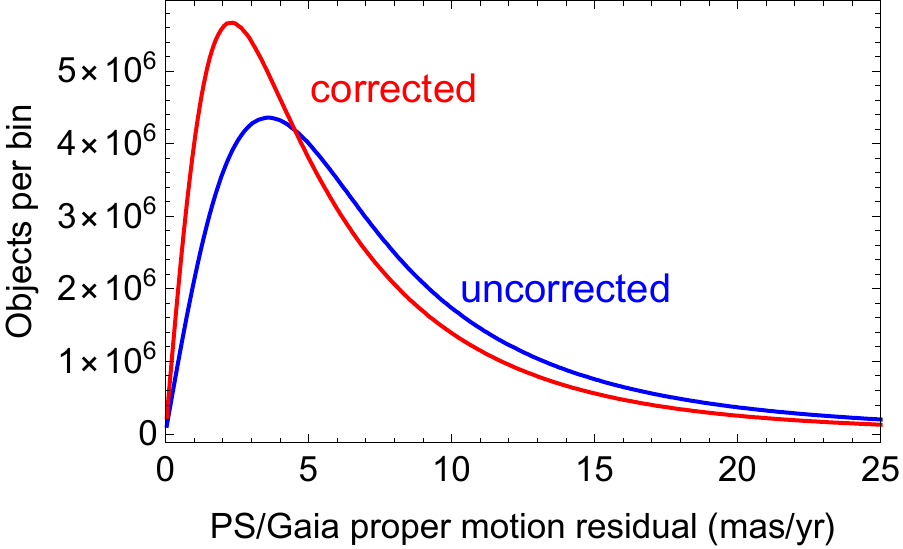}
\caption{Distribution of PS1/Gaia proper motion residuals in mas/yr with and without the PS1 corrections described in Section \ref{sec:algorithm} for PS1 reference objects. The bin size is 0.1 mas/yr.}
\label{fig:globalpm}
 \end{figure}

Figure~\ref{fig:globald} plots the distribution of positional residuals between PS1 and Gaia for reference objects with both the uncorrected and the corrected PS1 astrometry based
on the algorithm in Section \ref{sec:algorithm}. The uncorrected residuals result from the  differences between the uncorrected PS1 positions and Gaia  positions that are corrected
for proper motion and parallax, as described in Steps 3 and 4  in the algorithm of Section \ref{sec:algorithm}. The corrected residuals result from the position differences after the
position shifts are applied from the medians the nearest 33 neighbors, as discussed in Steps 5 and 6 of the correction algorithm.
The figure shows that the corrected positional residuals are significantly smaller than the uncorrected values. The median values for the corrected and uncorrected residuals are 9.0 and 13.4 mas, respectively. The results in the figure for the entire PS1 survey area
are similar to those
for the sample field in Figure~\ref{fig:distortion-example-hist}.
The median of the PS1 positional errors in Figure~\ref{fig:globald} are reduced by about 33\%. This level of improvement is similar to the amount gained from using the Gaia proper motions and parallaxes in correcting Gaia object positions to the PS1 epoch.
The mode (peak) values for the corrected and uncorrected residuals are 4.4 and 7.8 mas, respectively.

The distribution functions for residuals such as in Figure~\ref{fig:globald} involve
a rapid rise to a peak value  followed by a more extended decline. These properties
can be understood as follows. In a simple model, 
the error distribution is taken to be a two-dimensional Gaussian 
that is isotropic in RA and Dec with a standard deviation $\sigma$. The distribution is then
\begin{equation}
    dN(x, y) = \frac{1}{2 \pi \sigma^2} \exp{\left(-\frac{x^2+y^2}{2 \sigma^2} \right)} dx \, dy,
\end{equation}
where $x$ and $y$ are the residuals in RA and Dec
as directed lengths, respectively. We make a change of variables
from Cartesian coordinate $(x, y)$ to polar coordinates $(r, \theta)$ and integrate over $\theta$
to obtain 
\begin{equation}
    \frac{dN(r)}{dr} = \frac{r}{\sigma^2} \exp{\left(-\frac{r^2}{2 \sigma^2} \right)}.
\label{rayleigh}
\end{equation}
The distribution function $dN/dr$ is then the well known Rayleigh distribution. It is similar to the distribution function in Figure~\ref{fig:globald}.
However, the standard deviation $\sigma$ is not uniform across all sources. It depends on magnitude, exposure time, morphology (star or galaxy), etc.  Consider a slightly more complicated model in which $\sigma$ depends on source magnitude $m$. Equation (\ref{rayleigh}) then generalizes to
\begin{equation}
    \frac{dN(r)}{dr} = r \int dm \, \frac{\rho(m)}{\sigma^2(m)} \, \exp{\left(-\frac{r^2}{2 \sigma^2(m)} \right)},
\label{nonrayleigh}
\end{equation}
where $\rho(m)$ is the distribution of sources by magnitude. Equation (\ref{nonrayleigh}) is no longer
of the form of a Rayleigh function because the sum of Gaussian functions that appear in the integral is not a Gaussian.
The distribution $\rho(m)$  generally increases with $m$.
Since the variation in $\sigma(m)$ is not very large (see below for details), the
resulting distribution function should be somewhat similar to a Rayleigh distribution. 

The initial rise to the peak value for the corrected distribution in 
Figure~\ref{fig:globald} can be well fit to a Rayleigh distribution. The main tail of the corrected distribution from 20 mas to 40 mas is well fit to an exponential, while far into the tail (40 to 100 mas) the distribution follows a power law with index $\sim -2.8$. Similar properties hold for the uncorrected distribution.

Figure~\ref{fig:globalpm} plots the distribution of the proper motion residuals  between the cross-matched PS1 reference objects and Gaia objects  with and without the corrections to the PS1  reference objects. In this case, the corrections provide a smaller improvement to the residuals.  The median values for the corrected and uncorrected residuals are 4.8 and 6.3 mas, respectively. The median of the proper motion errors are therefore reduced by about 24\%.
The mode (peak) values for the corrected and uncorrected residuals are 2.3 and 3.7 mas/yr, respectively.

\begin{figure}
\centering
	\includegraphics[width=0.82\columnwidth]{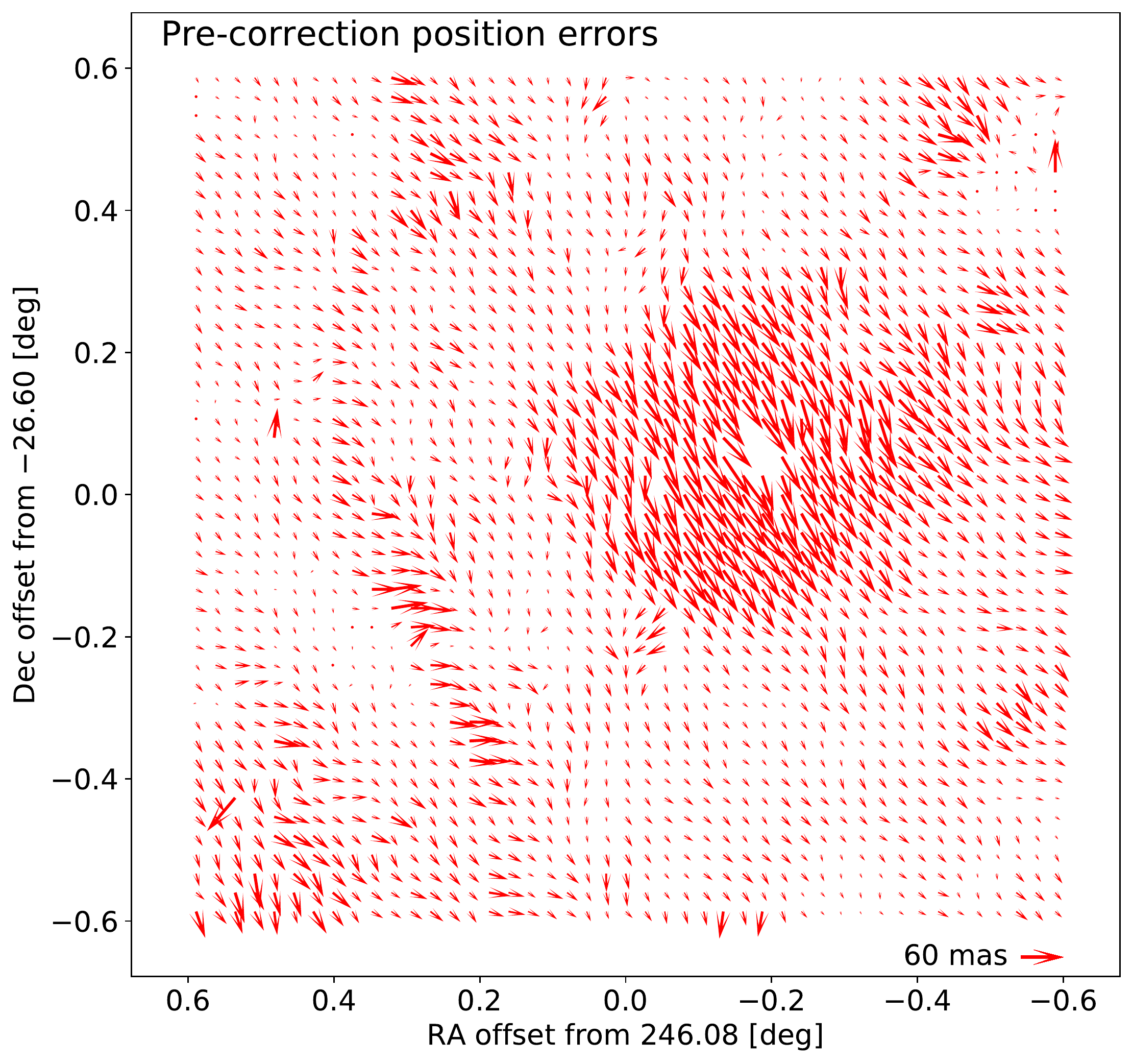}\hfil
	\includegraphics[width=0.82\columnwidth]{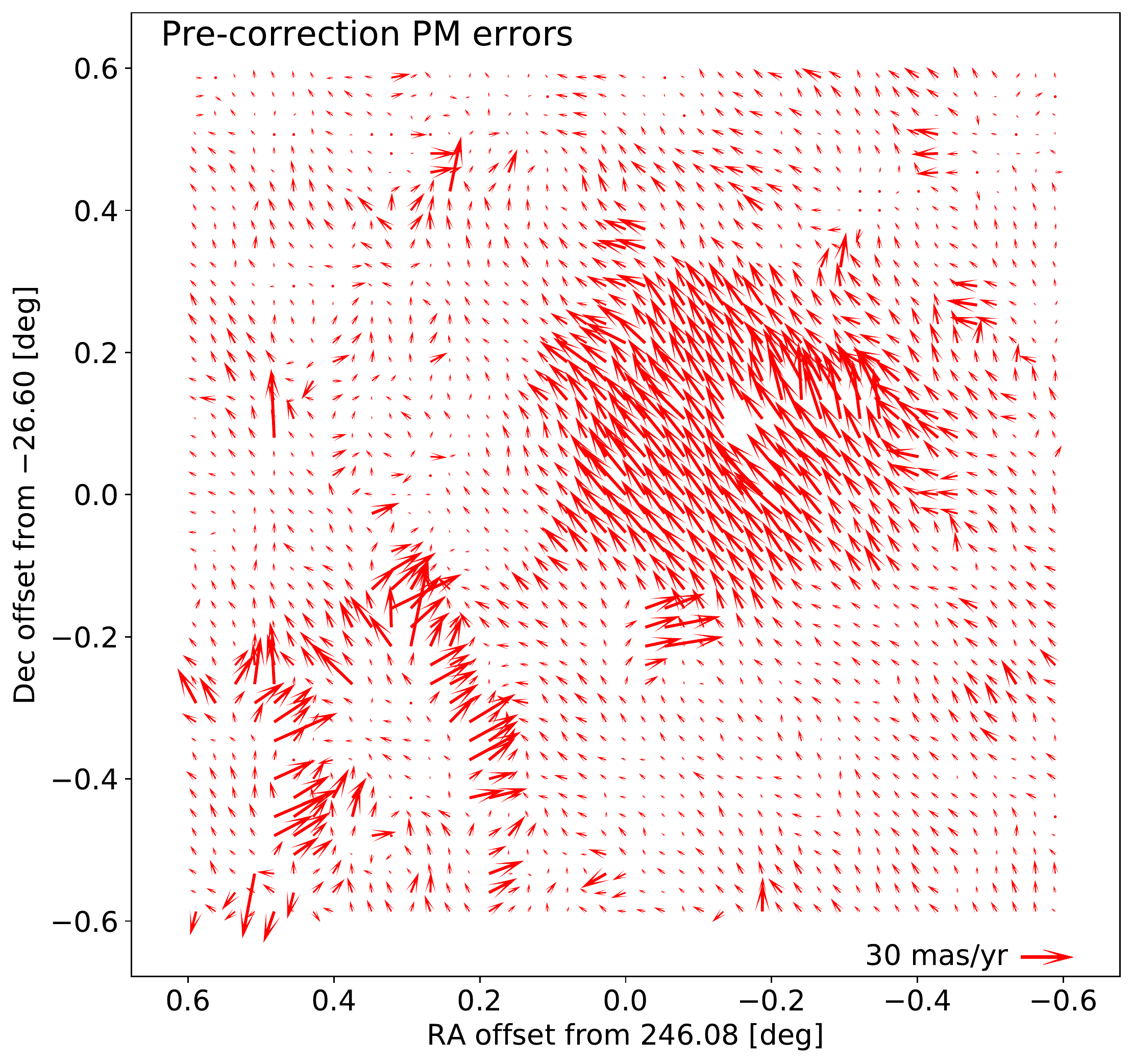}\hfil
	\includegraphics[width=0.82\columnwidth]{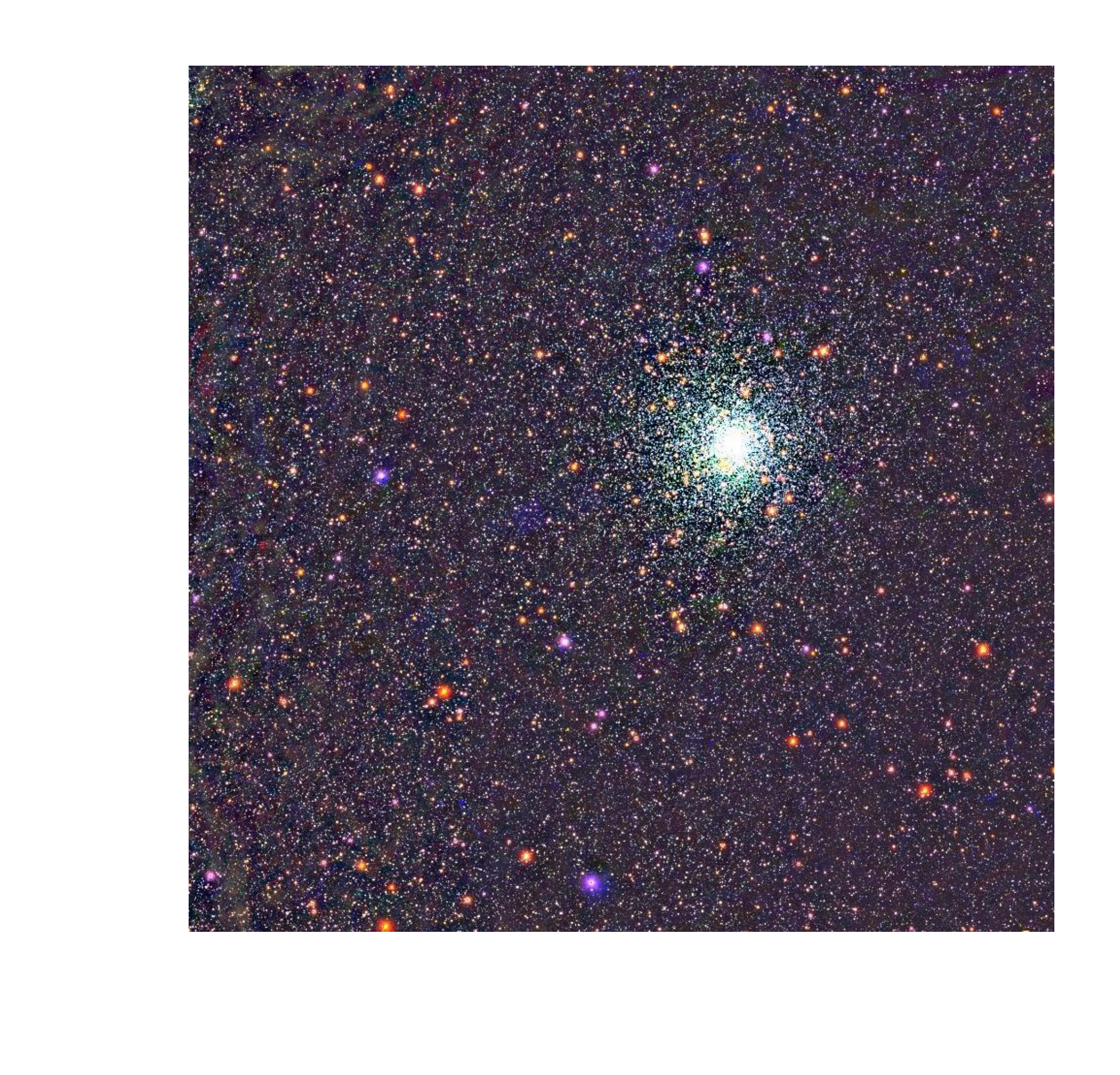}
	\caption{Distortions in position (top) and PM (center) in the PS1 catalog in the field of
	the globular cluster M4.  The median offsets between PS1 and Gaia matches are computed in spatial cells
	as in Fig.~\ref{fig:distortion-example}.  Large systematic errors are seen in both position
	and PM, with distortions strongly correlated with the position of the globular
	cluster (bottom panel). The adaptive algorithm in this paper removes these distortions. }
\label{fig:m4}
\end{figure}

\begin{figure}
\centering
\includegraphics[width=0.95\columnwidth]{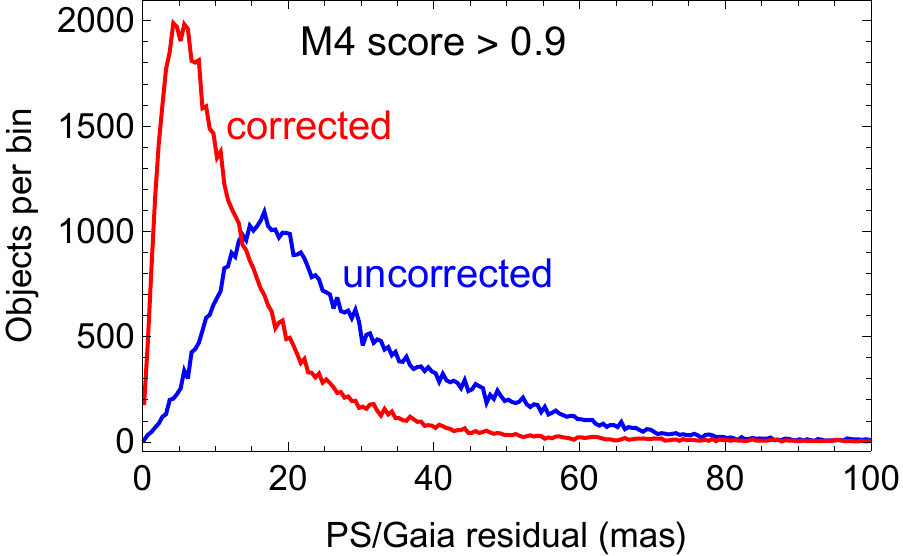}\hfil
\includegraphics[width=0.95\columnwidth]{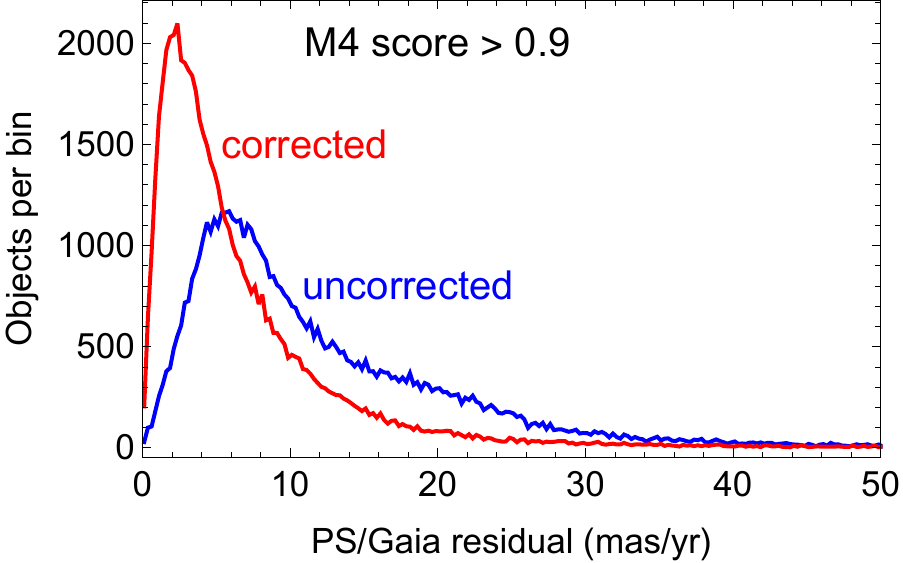}\hfil
	\caption{Histograms of the position (top) and PM (bottom) PS1/Gaia
	residuals before
	and after correction for the PS1/Gaia matches near globular cluster M4
	(Fig.~\ref{fig:m4}).  Both quantities are greatly improved by the removal
	of systematic distortions. The bin sizes are 0.5 mas and 0.25 mas/yr for the top and bottom plots, respectively}
\label{fig:m4-hist}
\end{figure}

While most fields have a relatively modest improvement in proper motion
accuracy from our processing, there are some dramatic exceptions.
Figure~\ref{fig:m4} shows positional and proper motion errors in the vicinity
of globular cluster M4.  The PS1 image is also shown for reference.  The
globular cluster has a strong influence on the errors; the proper motion
errors in particular are affected by the cluster's large relative PM of about
18 mas/yr compared with the field stars \citep{Cu1990,Wa2018}.
We attribute the
errors to the cluster's PM rather than simple crowding because a plot of the
mean PM as a function
of sky position in this field (not shown) reveals that the cluster's PM compared with the field
has effectively been removed by the PS1 DR2 astrometric calibration.
The adaptive
correction algorithm leads to large improvements in both positions and PMs in
this field (Fig.~\ref{fig:m4-hist}).

\begin{figure}
\centering
    \plotone{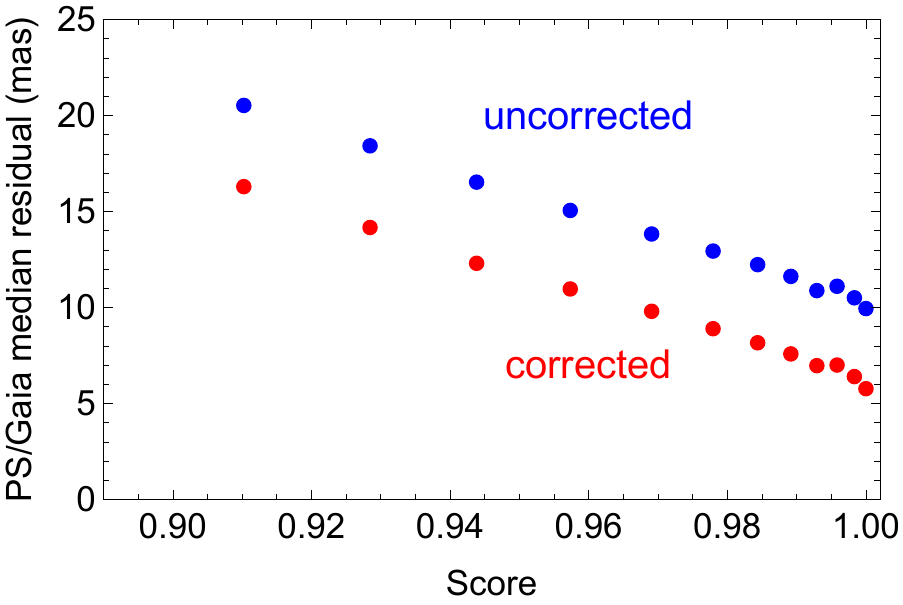}
\caption{The median PS1/Gaia positional residuals in mas as a function of \psscore\ for PS1 reference objects. The number of objects is the same between adjacent points.}
\label{fig:dscore}
 \end{figure}

We now examine some factors that influence the residuals.
Figure~\ref{fig:dscore} plots the positional residuals as a function of the \psscore\ for PS1 reference objects. The plot is made by constructing bins in intervals of \psscore\ values that contain an equal number of objects. For each bin we determine the abscissa or $x-$value of the plotted point as the  median \psscore\ within the bin. The ordinate or $y-$value is the median
value of the residual within the bin.
Similar plots are described in the remainder of this section.
Since there are an equal number of objects between adjacent
points, the concentration of points near a \psscore\ of 1 (the value for a point source) indicates that about half the reference objects have a \psscore\ greater than 0.98. Both the corrected and uncorrected residuals generally decline with increasing \psscore. For a \psscore\ close to 1, the corrected and uncorrected residuals are about 5.8 and 9.9 mas, respectively. The correction then provides about a 40\% improvement for the most point-like PS1 objects, which are less affected by blending in crowded regions and so ought to show better agreement with the Gaia positions.
For the lowest \psscores\ considered, close to 0.91,  the corrected and uncorrected residuals are about 16.3 and 20.5 mas, respectively, so the correction provides only about a 20\% improvement. Since many more objects have a \psscore\ closer to 1 than 0.91, the overall improvement is about 33\% as discussed above.

\begin{figure}
\centering
    \plotone{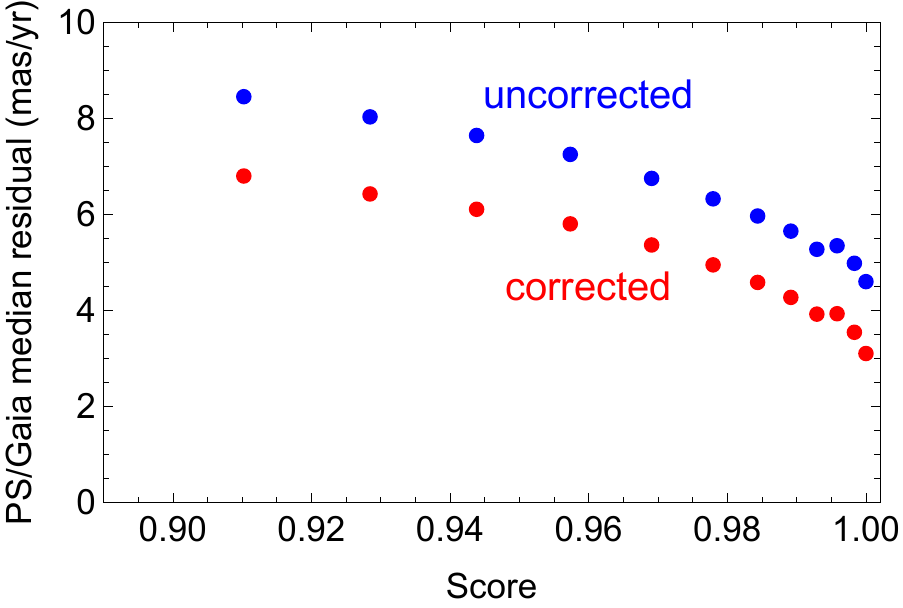}
\caption{The median PS1/Gaia proper motion residuals in mas/yr as a function of \psscore\ for PS1 reference objects.}
\label{fig:pmscore}
 \end{figure}

Figure~\ref{fig:pmscore} plots the proper motion residuals as a function of the \psscore\ for PS1 reference objects. In this case, we again find that the residuals
generally decrease with increasing \psscore. The corrected and uncorrected residuals
are about 6.8 and 8.5 mas/yr near a \psscore\ of 0.91, respectively and about 3.1 and 4.6 mas/yr
near a \psscore\ of 1, respectively.

 \begin{figure}
\centering
    \plotone{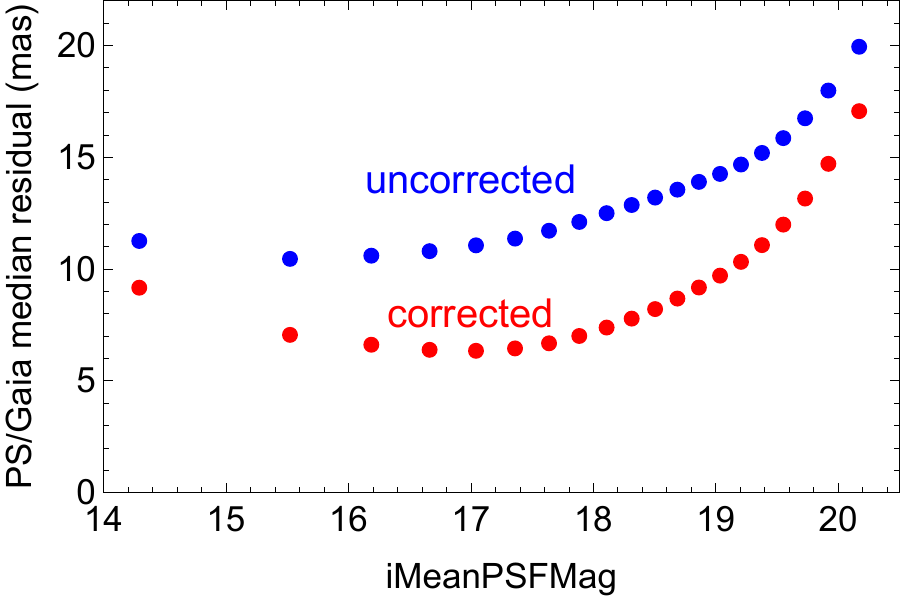}
\caption{The median PS1/Gaia position residuals in mas as a function of PS1 $i$-band PSF magnitude for PS1 reference objects.}
\label{fig:dimag}
 \end{figure}

\begin{figure}
\centering
    \plotone{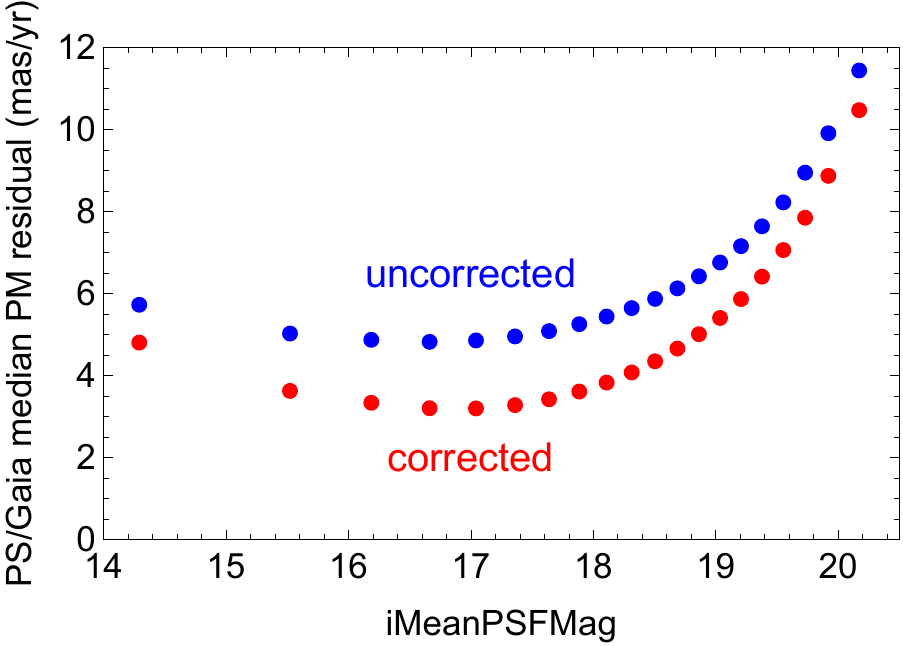}
\caption{The median PS1/Gaia proper motion residuals in mas/yr as a function of PS1 $i$-band PSF magnitude for PS1 reference objects.}
\label{fig:pmimag}
 \end{figure}

Figure~\ref{fig:dimag} plots the positional residuals as a function of the PS1 $i$-band PSF magnitude found in the PS1 \textit{MeanObject} table (\texttt{iMeanPSFMag}). Both the corrected and uncorrected residuals increase at bright and faint magnitudes.  The corrected residuals reach a minimum at intermediate values of about 17 mag, where the corrected and uncorrected residuals are about 6 and 11 mas, respectively. The faintest magnitude plotted of about $i=20.5$ is close to the Gaia magnitude limit. The residual increase at the bright end is likely due to effects of saturation
(for the brightest stars) and to the Koppenh\"ofer Effect \citep{Magnier2016}, which generates brightness-dependent position errors in the PS1 detectors. The increase at
the faint end is due to the decreasing signal-to-noise ratio in the PS1 measurements (and to a lesser extent
in the Gaia measurements as well).
The corrected residuals have a larger variation with magnitude than
the uncorrected residuals. At the brightest and faintest magnitudes, the corrected and uncorrected residuals are about the same because
the errors are dominated by effects other than the astrometric calibration.
Similar patterns are found for other PS1 magnitudes.

Figure~\ref{fig:pmimag} plots the proper motion residuals as a function of the PS1 $i$-band PSF magnitude.
In this case, we again find that the residuals have a minimum value at about 17 mag. The corrected residuals increase from the minimum residual to the largest at the faintest magnitudes by about a factor of three.

\section{Object Counts in Declination Stripes}
\label{sec:objprop}

 \begin{figure}
\centering
    \plotone{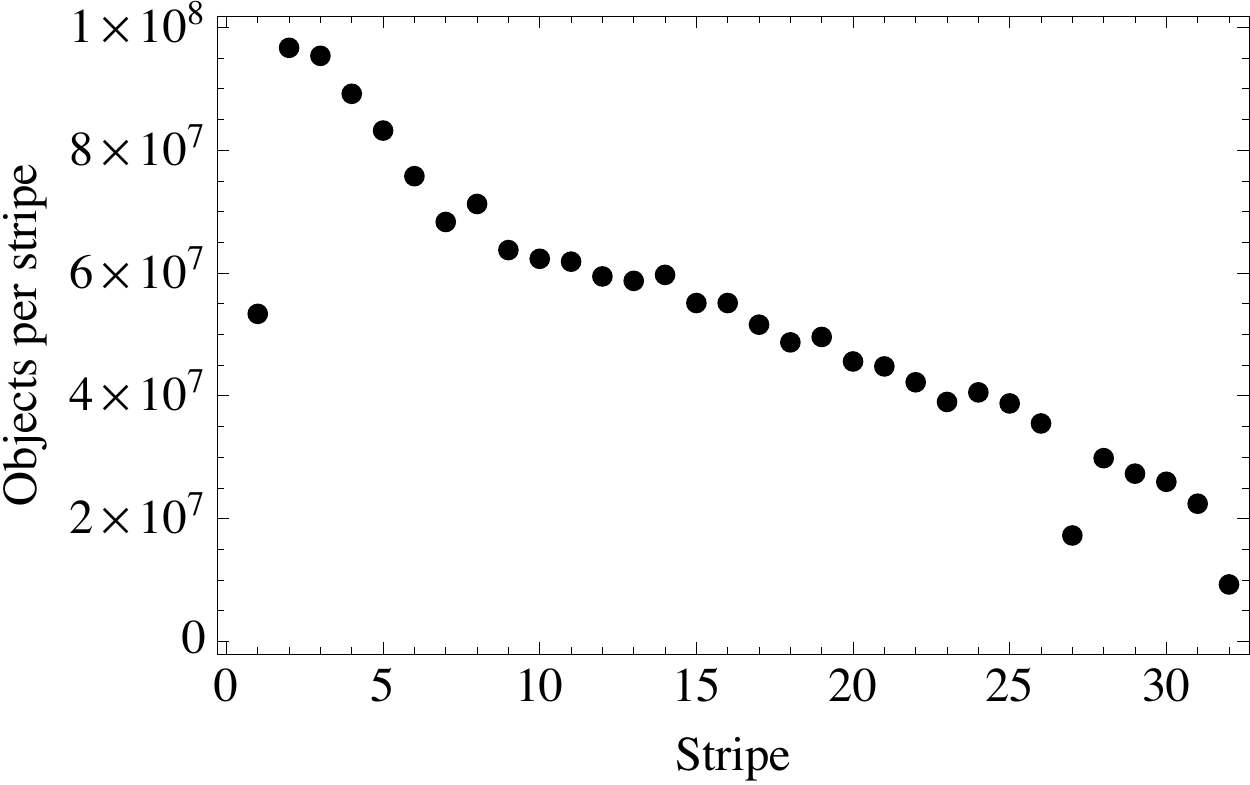}
\caption{Number of PS1 objects that we consider in each stripe.}
\label{fig:numobj}
 \end{figure}

The PS1 pipeline partitions the data into 32 stripes that are defined as
bands in declination that have a roughly equal number of detected objects.
We make use of this partitioning in our calculations.
The width of each stripe in declination varies
somewhat, but is typically about 3.3 degrees.
The stripes are labeled by integers 1 to 32 that increase with declination.
Table \ref{tab:stripes} in  Appendix \ref{sec:Stripes} contains
the declination ranges in each stripe.
Figure~\ref{fig:numobj} plots the number of all PS1 objects that we consider for astrometric correction in each stripe.
The number of objects generally decreases with the stripe number
or declination. Notice that Stripes 1, 27, and 32 are anomalous
compared to neighboring stripes. The number of objects in Stripe 1 is low because of the cut off in declination at $\delta > {-}30^\circ$. Stripe
32 contains a small number of sources due to its northernmost position ($\delta > 77.4^\circ$).

We have found that most of the objects that should be in Stripe 27 are missing. This is the result of an
error in the database population for the \textit{Detection} table in the PS1 DR2 database.
In addition, the spatial distribution of missing objects is irregular which results in additional errors in the astrometric corrections that we undertake. Work is underway to fix this problem, but for now there is missing data in our corrected astrometry tables for
declinations between 54.2 and 57.6 degrees.

\begin{figure}
\centering
    \plotone{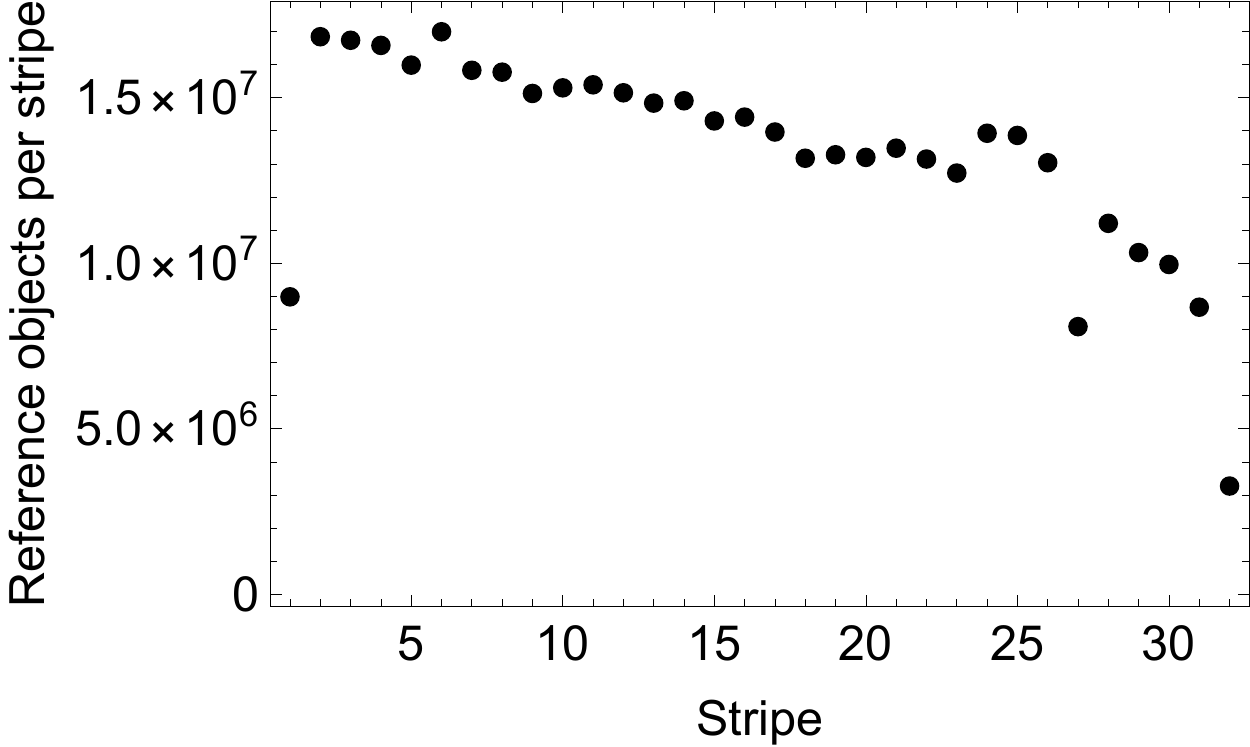}
\caption{Number of reference PS1 objects in each stripe.}
\label{fig:numref}
 \end{figure}

Figure~\ref{fig:numref} plots the number of reference PS1 objects in each stripe.
This distribution function is flatter than the distribution function of PS1
objects in Figure~\ref{fig:numobj}.  This is a consequence of the
much greater crowding in the Galactic plane near the Galactic center, which
results in a higher fraction of sources being rejected as reference objects by
the \psscore\ criterion due to blending with neighboring objects.

\begin{figure}
\centering
    \plotone{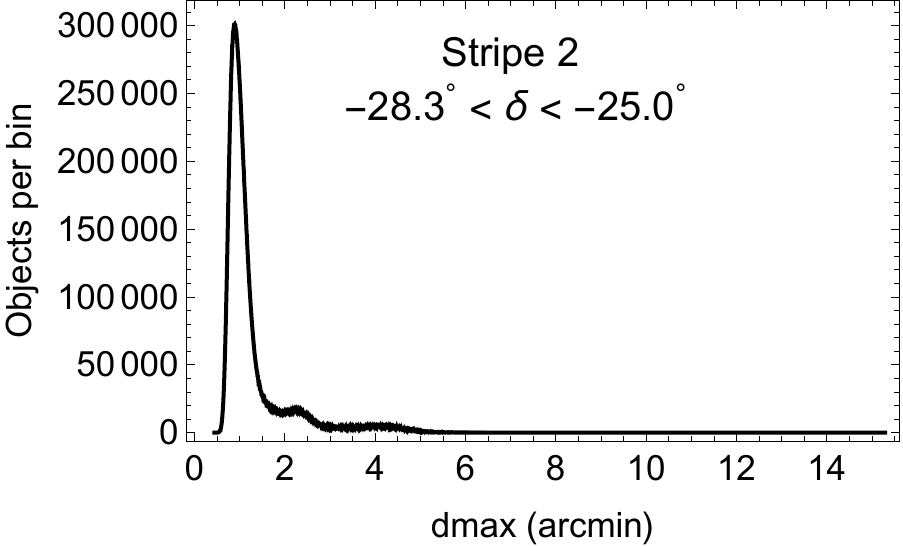}
\caption{Distribution of the maximum distance of the nearest 33 reference objects to each PS1 object in Stripe 2. The bin size is 0.1 arcsec. }
\label{fig:dmax}
 \end{figure}

In Step 5 of the astrometric correction algorithm,
we utilize the nearest 33 reference objects.
The maximum distance to the nearest 33 PS1 reference objects varies with object density. Figure~\ref{fig:dmax} is a histogram of this distance for
PS1 objects in Stripe 2. The maximum peak is near 1 arcmin. However, there are values that extend to as far as about 15 arcmin for a small fraction of sources in
low density regions of reference objects.
There are a variety of reasons for a very low density of reference objects.
It can be simply due to a low density of general PS1 objects, but also can be due to a low density of PS1 objects with acceptable \psscore\ values, a
low density of Gaia objects with proper motions, etc.
Structures in the residuals on scales larger than $\sim1$ arcmin are then smoothed by the correction algorithm.

\section{Position Residuals of Reference Objects in Declination Stripes}
\label{sec:posresw}

\begin{figure*}
\centering
\includegraphics[width=\textwidth]{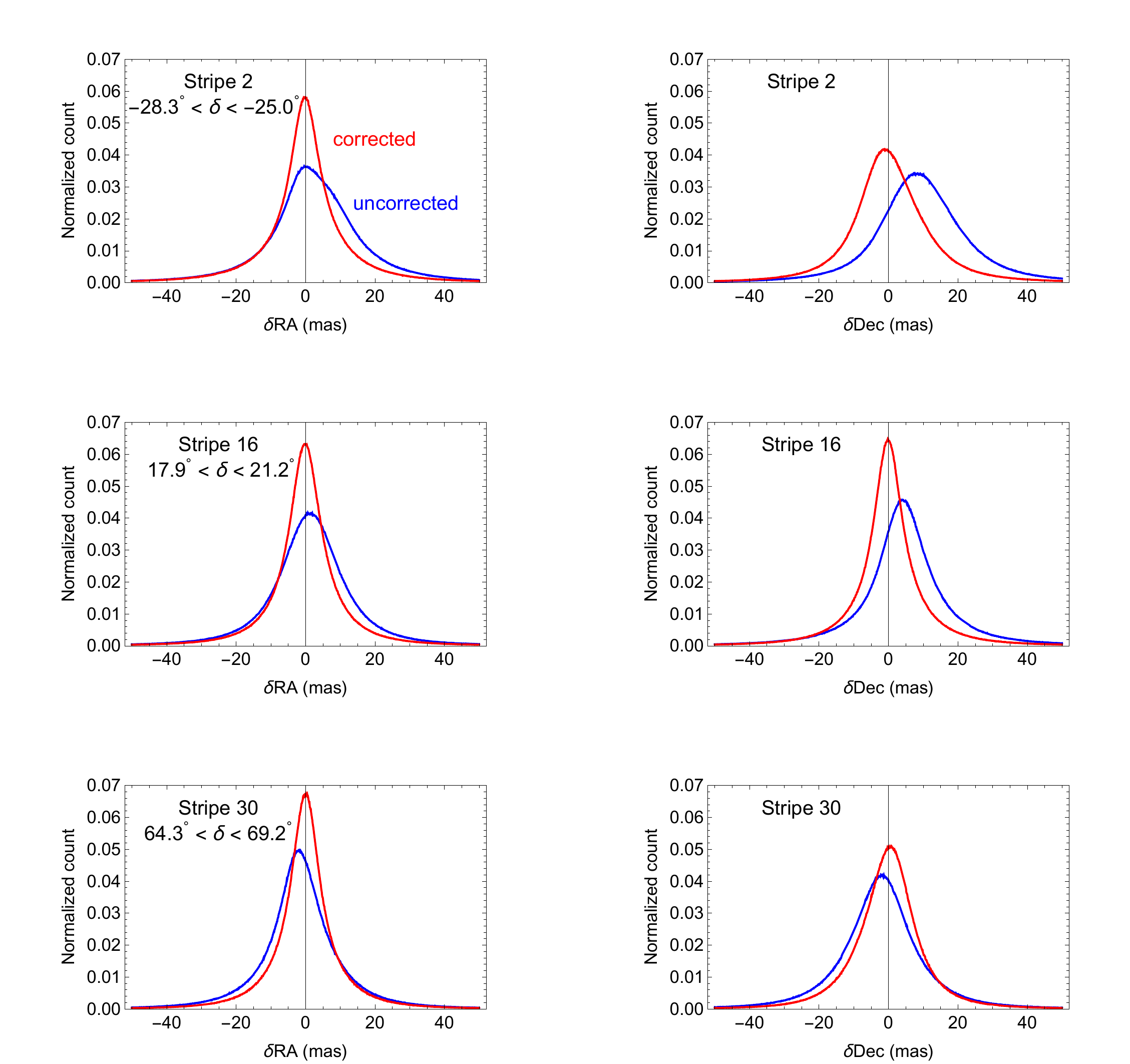}
\caption{Normalized count distribution of the PS1 to Gaia position residuals in three different stripes that cover the indicated declination ranges. The left column is for  $\delta$RA and the right for $\delta$Dec. The blue lines are for the uncorrected residuals and the red lines for the corrected residuals.  The normalization is such that the integral under each curve is unity.}
\label{fig:resstripes}
 \end{figure*}

In Section \ref{sec:global} we described the distribution of the PS1/Gaia reference object residuals globally, across all declinations. In this section, we consider how these residuals vary by declination stripe and consider residuals in RA and declination separately.
 We denote the position residuals in RA and declination by $\delta$RA and $\delta$Dec, respectively.
The residual for an object is the Gaia source position  minus the cross-matched PS1 object position.

Recall that the uncorrected residuals are simply the offsets from the uncorrected PS1 positions to the proper motion and parallax corrected Gaia
positions. On the other hand, the corrected residuals do not involve the PS1 to Gaia shift of the object being corrected
in making the PS1 position correction. It instead uses the shifts of the 33 nearest other reference objects.

The results in Figure~\ref{fig:resstripes} show significant offsets of the peaks from zero residual for the uncorrected residuals.
Much smaller peak offsets occur for the corrected distributions for both $\delta$RA and $\delta$Dec. In addition, the corrected distributions are narrower
than the uncorrected cases.

\begin{figure}
\includegraphics[width=0.95\columnwidth]{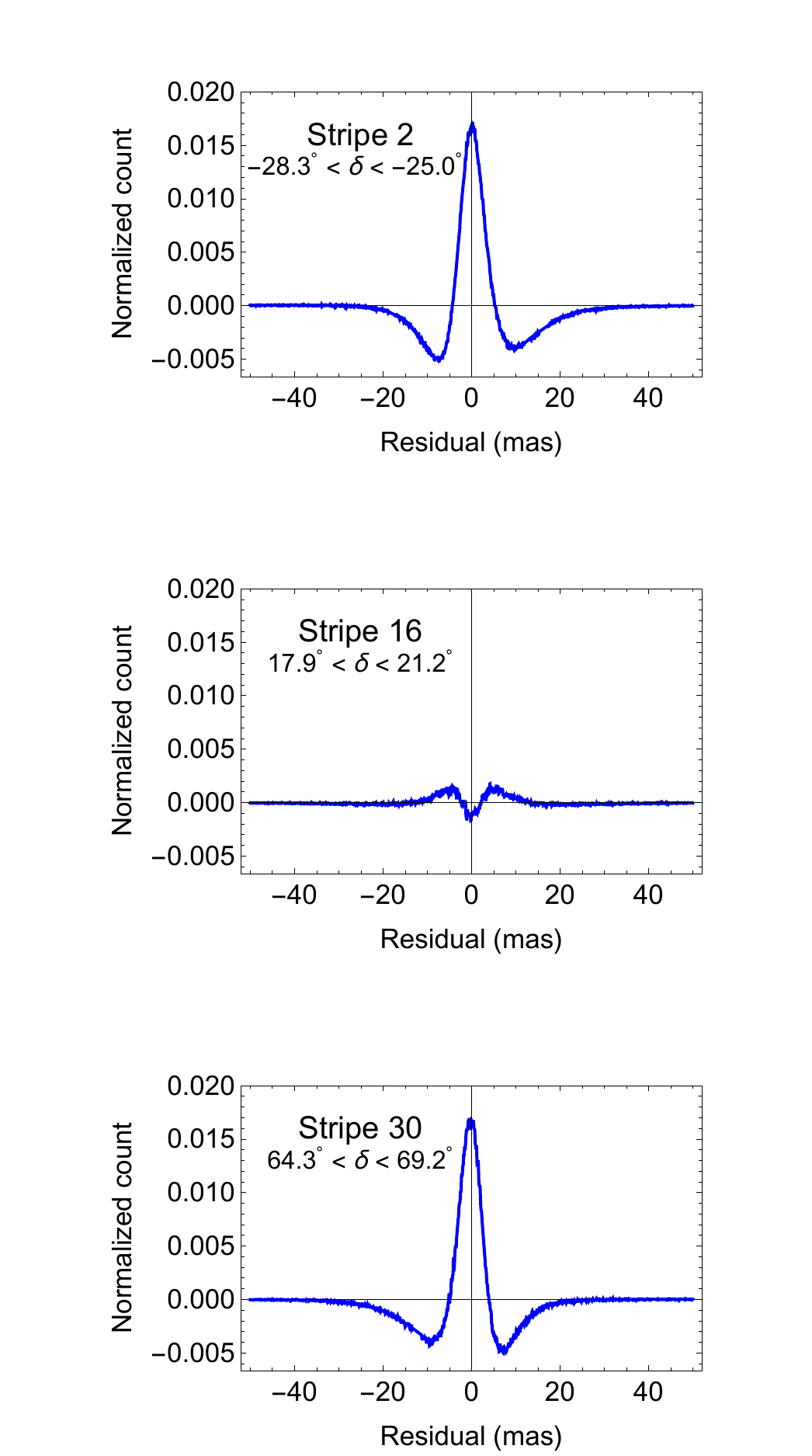}
 \caption{Difference in the corrected distributions of $\delta$RA and $\delta$Dec in Figure~\ref{fig:resstripes} for three stripes.  }
 \label{fig:ramdec}
 \end{figure}

We examine the difference between the RA and declination corrected residual distributions for a given stripe.
Figure~\ref{fig:ramdec} plots the RA residual distribution minus the declination residual distribution, i.e., the difference between the red curves in Figure~\ref{fig:resstripes} for each of the three stripes.
The curves for Stripes 2 and 30 are quite similar, with an RA distribution that is narrower than the declination distribution.  Note that both stripes are 47 degrees away from the zenith at the Haleakala Observatory, but on opposite sides of the sky. There is a slight asymmetry
in the distribution for Stripe 2 about zero residual
that is the mirror image of the asymmetry in Stripe 30.
The difference curve in Figure~\ref{fig:ramdec}
for Stripe 16 (which passes through the zenith) has
a much smaller variation, with an RA distribution that is slightly broader than the declination distribution. These results suggest that
the difference in the RA and declination residuals involves differential refraction effects from the Earth's atmosphere that are minimized at zenith angle zero.
For objects observed crossing the meridian, differential refraction effects primarily affect the declination.  In addition, for objects observed crossing the meridian, refraction effects should be the same for Stripes 2 and 30, which are equidistant from the zenith. The slight mirror image antisymmetry between the distributions in Stripes 2 and 30 is likely a consequence of some object observations being somewhat off the meridian in RA.

As in Section \ref{sec:global}, we see that the corrections significantly improve the agreement with Gaia astrometry. As we will see next, the residuals
involve contributions that are independent of declination, as found for Stripe 16, as well as the declination-dependent residuals as found for Stripes 2 and 30.
We examine the properties of the corrected PS1 to Gaia position residuals of the reference objects in each stripe.

\begin{figure}
\centering
\includegraphics[width=0.95\columnwidth]{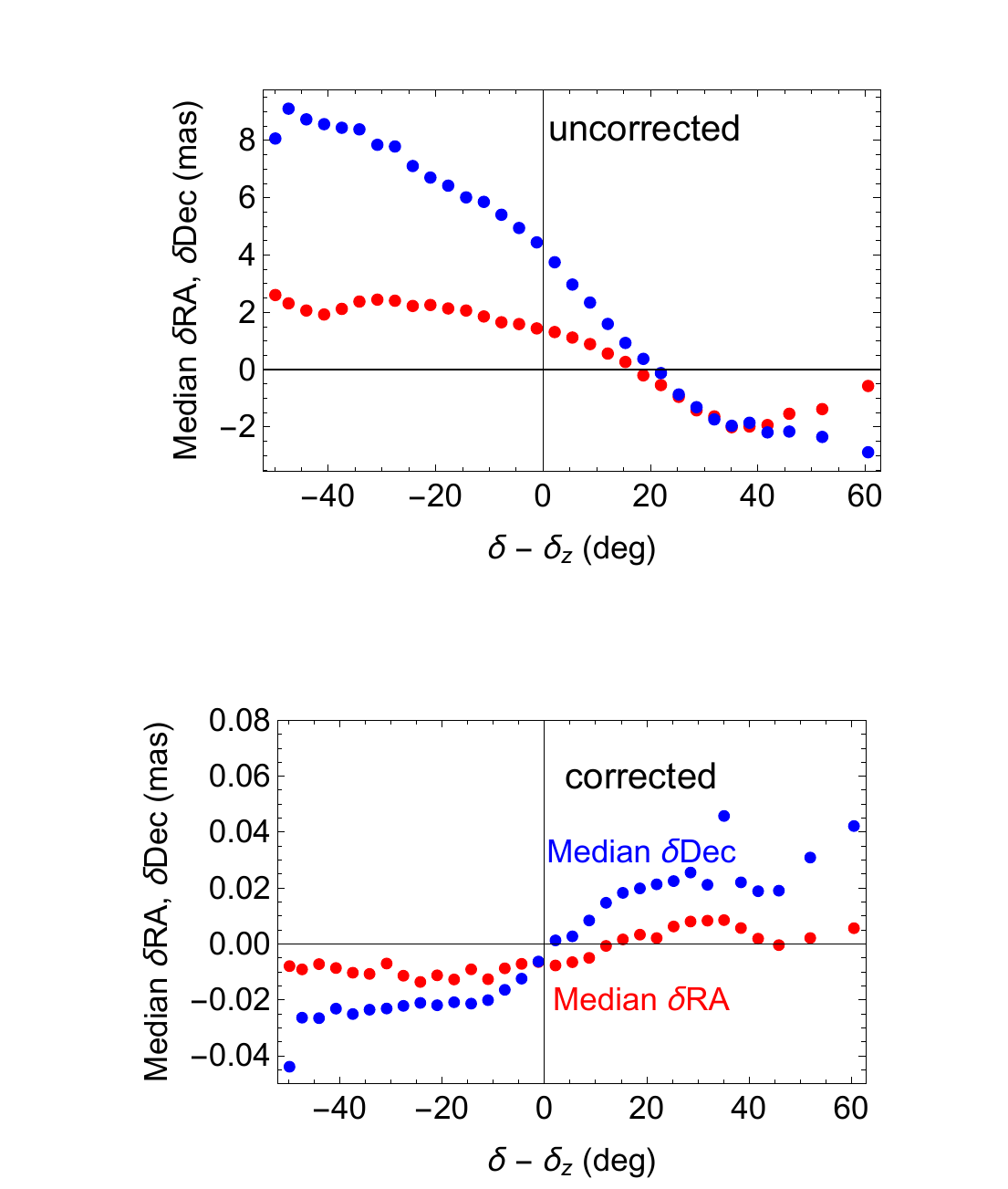}
\caption{Median of the position residuals in each stripe as a function of the difference of the average declination in the stripe $\delta$ from the declination of the zenith $\delta_{\rm z}$, with one point per stripe. Note that the $y$-axis range is much smaller in the
bottom panel.}
\label{fig:resmed}
 \end{figure}

\begin{figure}
\centering
\includegraphics[width=0.95\columnwidth]{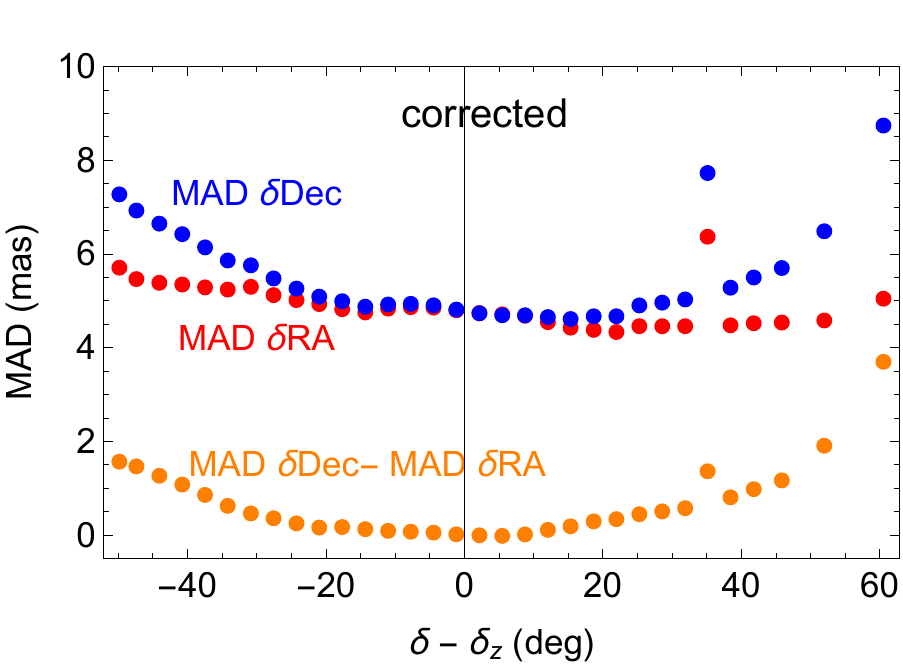}
\caption{Median absolute deviation (MAD) of the position residuals in each stripe as a function of the difference of average declination in the stripe $\delta$ from the declination of the zenith $\delta_{\rm z}$.  }
\label{fig:resmad}
 \end{figure}

Figure~\ref{fig:resmed} plots the systematic errors in the positions before and after correction. A point for each stripe shows the median of the coordinate residuals as a function of the difference of the average declination $\delta$ in the stripe from the declination of the zenith $\delta_{\rm z}$. Note that $|\delta - \delta_{\rm z}|$ is the average zenith angle in the stripe. The median of the uncorrected residuals is of order 5 mas. The medians of the corrected residuals are less than 0.1 mas, about two orders of magnitude smaller than the uncorrected residuals.
Notice that the corrected values generally increase with $\delta - \delta_{\rm z}$.
The declination residuals are smallest near Stripe 16, which passes through the zenith. This result is again consistent with the idea that the residuals are partly due to the effects of atmospheric refraction.

Figure~\ref{fig:resmad} plots the median absolute deviation (MAD) of the corrected coordinate residuals as a function of  $\delta - \delta_{\rm z}$ for each stripe.
 The MAD at zenith angle zero is about 4.8 mas, roughly corresponding to a standard deviation of about 7 mas for both $\delta$RA and $\delta$Dec. As expected from the earlier plots, the distribution is roughly symmetrical about zero zenith angle
 ($\delta=\delta_{\rm z}$). The MAD of $\delta$Dec
generally increases away from the zenith angle zero, while the MAD of $\delta$RA generally decreases slowly with increasing declination. The difference in MAD between $\delta$RA and $\delta$Dec is typically of order 1 mas. Notice that Stripe 27 at a declination of $\delta \sim 35^{\circ}$ shows larger errors than
expected, which are the result of the smaller number of PS1 objects detected and the irregular spatial distribution of objects in this stripe (see Section \ref{sec:objprop}).

\begin{figure}
\centering
    \plotone{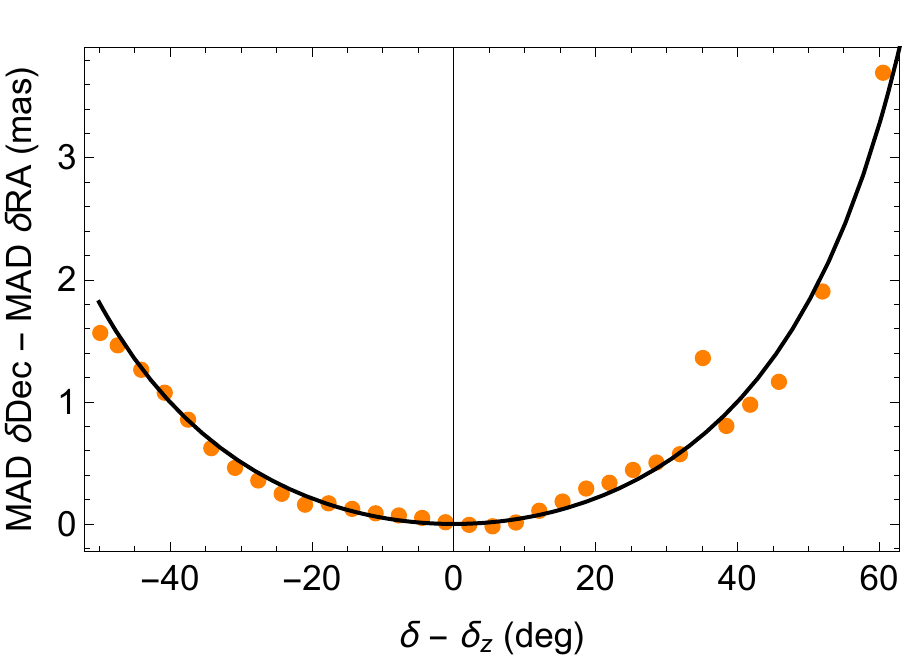}
\caption{The dots are the MAD of $\delta$Dec minus MAD of $\delta$RA in the corrected case for each stripe as function of the difference of average declination in the stripe $\delta$ from the declination of the zenith $\delta_{\rm z}$ (same as the lowest set of points in the bottom panel panel of Fig.~\ref{fig:resmad}). The solid line is given by Equation (\ref{eq:f1}) that is proportional to the increase in air mass away from the zenith direction. }
\label{fig:resmada}
 \end{figure}

We further examine the properties of the MAD difference between $\delta$RA and $\delta$Dec for corrected shifts. Figure~\ref{fig:resmada} plots the MAD of
 $\delta$Dec minus the MAD of
 $\delta$RA versus $\delta - \delta_{\rm z}$ for each stripe. The solid line is given by
 \begin{equation}
     f(z) = 3.27\, {\rm mas} \,(\sec(z) -1),
     \label{eq:f1}
 \end{equation}
 where
 \begin{equation}
 z= \delta - \delta_{\rm z}.
 \label{eq:z}
 \end{equation}
 Function $f(z)$ is proportional to the increase in air mass away from the zenith.

In a simple model for a single color, the deviation angle caused by refraction is given by
$ \sim 1 \, {\rm arcmin} \, \tan(z)$.
Consequently, we expect the MAD of the declination due to refraction
to be of order $ 1 \, {\rm arcmin} \sec^2(z) \, \delta z/4 \sim  1\, {\rm arcsec} \sec^2(z)$,  where $\delta z$
is the angular width of a stripe.
The magnitude of the effect that we find in Figure~\ref{fig:resmada}
is much smaller.
So clearly a large correction has already been made in the PS1 pipeline.

This analytic fit further supports the idea the difference in the RA and declination residuals involves the effects of refraction in the Earth's atmosphere. Such effects should exhibit a color dependence, which
we describe next.

\section{Color Effects}
\label{sec:color}

We examine the effects of color on the difference between the  corrected $\delta$RA and $\delta$Dec distributions. Color corrections to PS1 astrometry due to differential chromatic refraction (DCR) have been made in the PS1 pipeline,
as described in section 6.1.2 of \cite{Magnier2016}.  However, there are small color-dependent effects remaining after
our astrometric corrections.
We again consider
 PS1 objects that are cross-matched to Gaia and use the color information obtained from the Gaia catalog in cases where the color is determined. We define the color as the
magnitude in the Gaia blue passband minus the magnitude in the Gaia red passband.

\begin{figure}
\centering
\includegraphics[width=0.95\columnwidth]{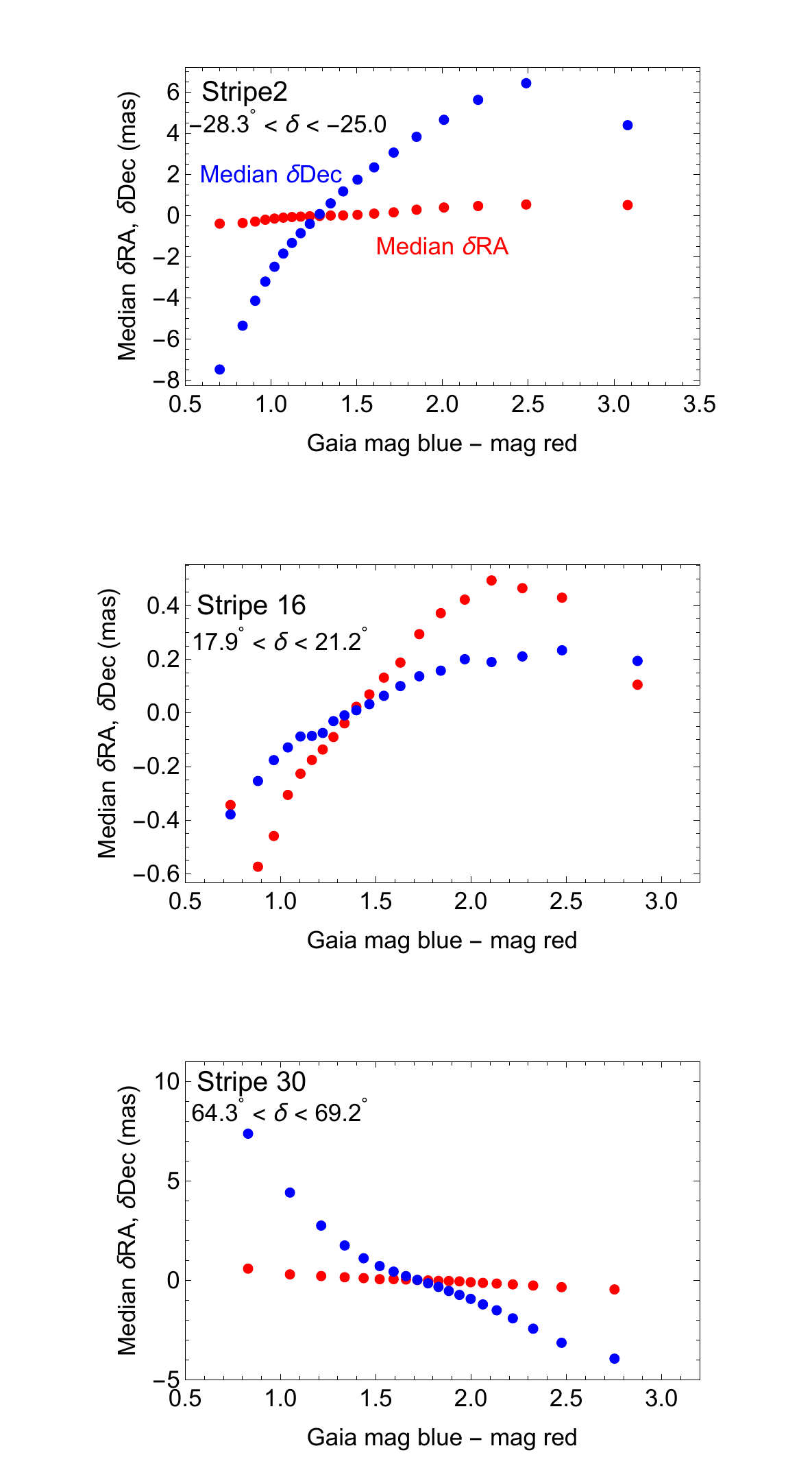}
\caption{Plot of median PS1 to Gaia position residuals for  corrected reference objects as a function of Gaia color for three stripes.  }
\label{fig:colormed}
 \end{figure}

Figure~\ref{fig:colormed} plots the median $\delta$RA and $\delta$Dec values for the three stripes as a function of Gaia color. Similar to Figure~\ref{fig:dscore}, the points plotted are binned such that there
are an equal number of PS1 reference objects between successive points.  The median residuals for Stripe 2 generally increase from blue (smaller color differences) to red (larger color differences). The
residual variation of $\delta$Dec with color is much larger than for $\delta$RA. A $\delta$Dec increase with color is expected due to the atmospheric
refraction of objects observed at declinations that are below the zenith declination ($\delta < \delta_{\rm z}$), as is the case here.
The small shift in $\delta$RA is also expected for objects observed close to but not on the
meridian.

It is not clear why $\delta$Dec decreases with color for the reddest bin in Stripe 2, although it could be related to differences in the
colors of stars in crowded regions of the Galactic plane. In any case, this decrease involves less than about 5$\%$ of the reference objects.

If we consider only objects with colors within 
one standard deviation of the median color (about 68\% of the bins),
the variation in $\delta$Dec with color is about 4~mas. This value roughly agrees with the approximately 3 mas standard deviation of inferred from the MAD of $\delta {\rm Dec}$ - MAD of $\delta {\rm RA}$ for this stripe in Figure~\ref{fig:resmad}. This variation
of $\delta$Dec with color largely explains why the $\delta$Dec distribution is broader than the $\delta$RA distribution in Figures~\ref{fig:resstripes} and \ref{fig:resmad}.

\begin{figure}
\centering
\includegraphics[width=0.95\columnwidth]{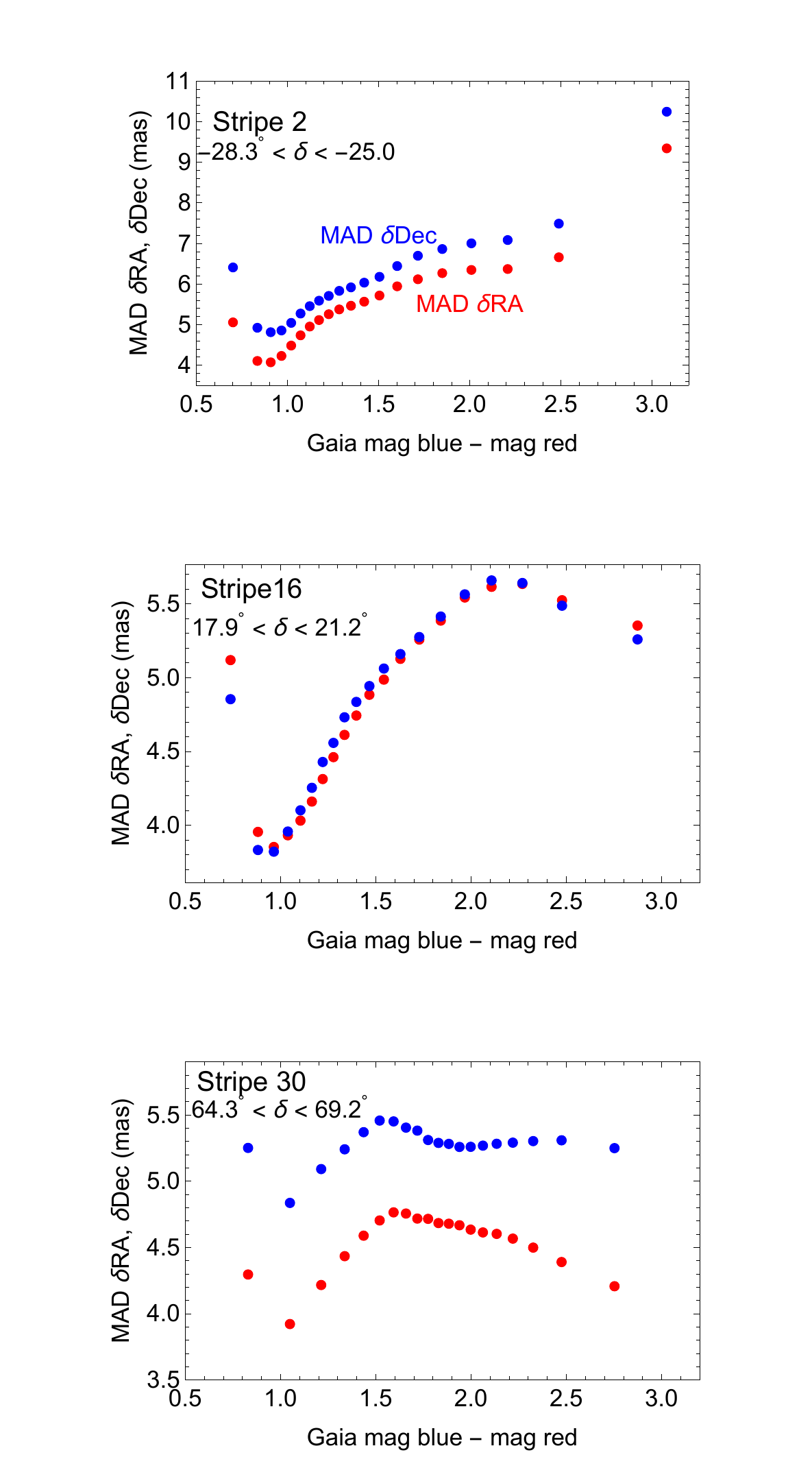}
\caption{Plot of median absolute deviations (MAD) of the PS1 to Gaia position residual for corrected reference objects as a function of Gaia color for three declination stripes. }
\label{fig:colormad}
 \end{figure}

The top panel of Figure~\ref{fig:colormad} shows the color dependence of the MAD in Stripe 2. As expected, the MAD for $\delta$Dec is larger than for $\delta$RA.
The MADs of $\delta$RA is smaller than the MADs of $\delta$Dec by a roughly constant amount.

Stripe 16 contains zenith angle zero. In this case
in Figure~\ref{fig:colormed}, the variation of the median $\delta$Dec with color is
much smaller than in the Stripe 2 case and smaller than the variation of the median of $\delta$RA in this stripe. This result is also consistent with the idea
that atmospheric refraction is the cause of the asymmetry.
There should be little differential refraction in the declination direction because objects in this stripe
are observed near the zenith with offsets in the RA direction. (The design of the
PS1 telescope mount prevents it from tracking objects closer to the zenith than 10--20 degrees.)
There is little difference between the MAD of $\delta$RA and the MAD of $\delta$Dec as a function of color in Stripe 16.

Consider Stripe 30 that has the same zenith angle as Stripe 2 but at higher declination ($\delta > \delta_{\rm z}$). In this case, the 
$\delta$Dec residuals plotted in the bottom panel of Figure~\ref{fig:colormed}
\textit{decrease} with color. This decrease is by a similar amount as the increase in the case of Stripe 2.
The sign difference in the variation again is consistent with expectations from atmospheric
refraction, where the natural direction is toward the zenith, meaning toward larger declinations in
Stripe 2 and toward smaller declinations in Stripe 30.
Again $\delta$RA residuals are expected to be small for objects observed
close to the local meridian.
As in the case of Stripe 2, the MADs of $\delta$RA and the MADs of $\delta$Dec for Stripe 30 are offset from each other by a roughly constant amount.

\begin{figure}[b]
\centering
   \plotone{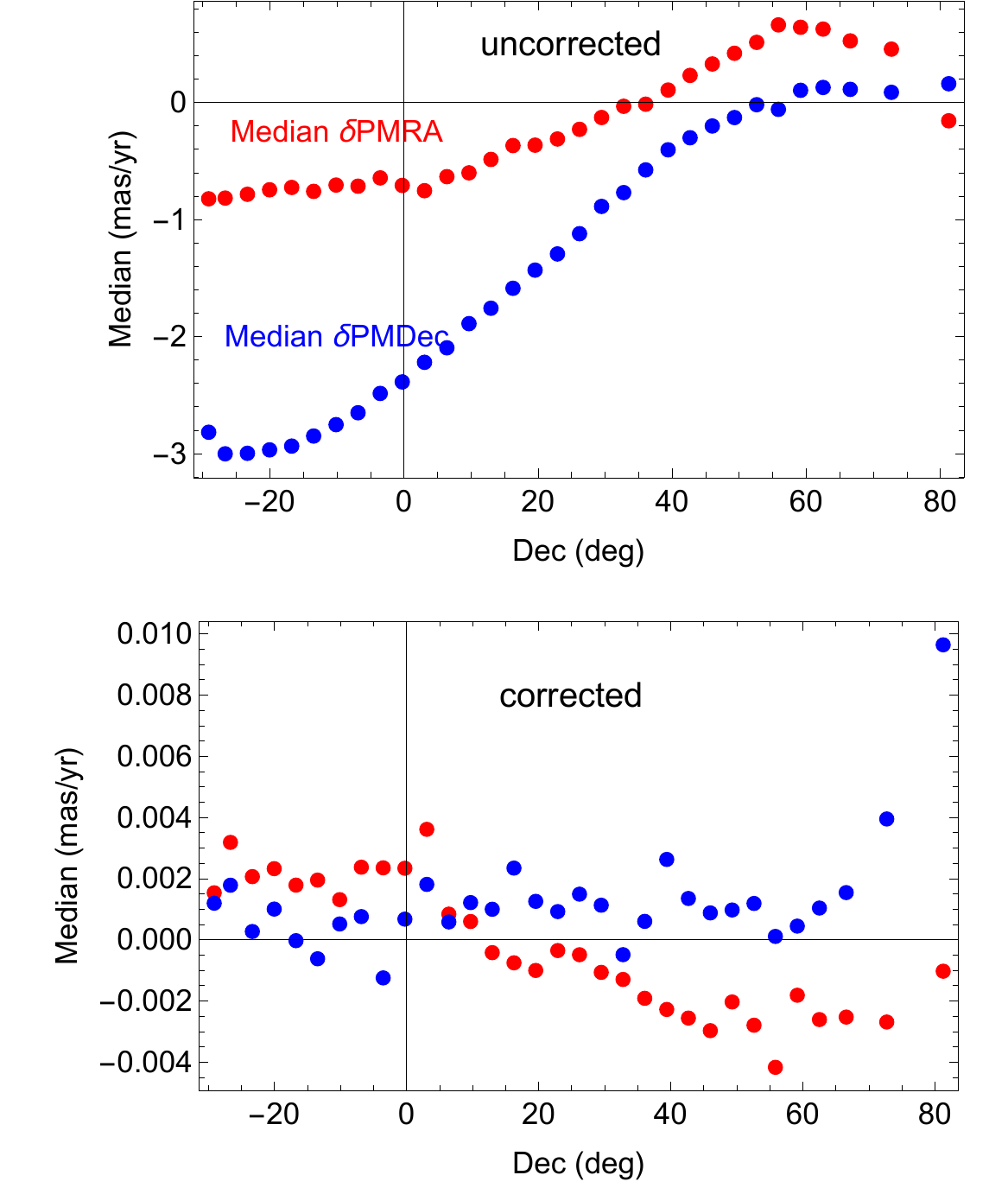}
\caption{Median  of  the  PS1  to  Gaia  proper motion  residuals as a function of declination with one point per stripe. The upper panel is for uncorrected proper motion values and the bottom panel is for our corrected proper motions.  }
\label{fig:respmmeddec}
\end{figure}

Atmospheric refraction affects PS1 astrometry but not Gaia astrometry.
We crudely approximate the PS1 filter throughput distribution
given in \cite{Tonry2012}
as a flat function between 550 nm and 900 nm.
The declination residuals expected by a wavelength change from  550 nm
to 900 nm  due to atmospheric refraction
is estimated by table 11.24 of \cite{Schubert2000}
as
\begin{equation}
   \delta {\rm Dec} \sim \delta R \, \tan(\delta-\delta_z)
   \label{eq:dDec}
\end{equation}
with $\delta R \sim 0.7$ arcsec.
If we take into account the variations of stellar spectral energy distributions in the PS1
passbands for a range of typical stellar temperatures of 4000 to 5000 degrees, then
the range of $\delta$Dec values is reduced so that
$\delta R \sim 0.026$ arcsec.
The $|\delta$Dec$|$ values
we find in Stripes 2 and 30, where $|\tan(\delta-\delta_z)| \simeq 1$,
are about a factor of 4 smaller than this crudely estimated value.
Recall that DCR corrections have already been applied to the PS1 DR2 catalog positions
by \cite{Magnier2016}.  The DCR effects in Figure~\ref{fig:colormed} are therefore smaller
amplitude variations that remain after the catalog corrections have been combined with the
corrections in this paper.

Figure~\ref{fig:colormed}
exhibits the qualitative dependence of the form given by Equation (\ref{eq:dDec}).
Stripe 16 of course has a small
$\delta$Dec because $\delta-\delta_z \simeq 0$.
In Stripe 2, $\delta$Dec increases with color, which implies a positive but small $\delta R$ value.
Stripe 30 also has the expected dependence with a similarly small
positive $\delta R$ value.

\section{Proper Motion Residuals in Declination Stripes}
\label{sec:pmresidstr}

A similar approach can be used to analyze the accuracy of our corrected proper motions compared with the Gaia PMs.
We again consider reference PS1 objects to examine their residuals from the corresponding Gaia sources, defined
as Gaia proper motion minus PS1 proper motion.
Figure~\ref{fig:respmmeddec} plots the median of the proper motion residuals in the uncorrected and corrected cases in each stripe. The corrected case reduces the median proper motion values from $\sim 2$ mas/yr to $\sim 0$ mas/yr.  The MAD of the proper motion residual values are nearly constant with stripe number, nearly the same for RA and Dec,
 and are $\sim 3.5$ mas/yr for the uncorrected case and  $\sim 2.5$ mas/yr in the corrected case.

Note that we do not expect to see significant effects from differential chromatic diffraction in the proper motions because
the stellar colors and airmass are usually very similar for observations of the same star at different epochs.

\section{Additional Checks}
\label{sec:checks}

In this section we describe some addition checks on the
astrometric improvements.

\subsection{Continuity of Residuals Across \psscore=0.9}
\label{sec:rescont}

The astrometric correction algorithm described in Section \ref{sec:algorithm} is based on the properties of a subset of objects called reference objects that are cross-matched
with Gaia DR2. The correction algorithm
relies on taking the median PS1/Gaia shift of the nearest 33 reference objects objects, excluding the object being corrected. We argued in the beginning of Section \ref{sec:global} that since these
corrections do not involve the position shift of the object being corrected, their residuals are a valid measure of the astrometric error.

\begin{figure}
\centering
    \plotone{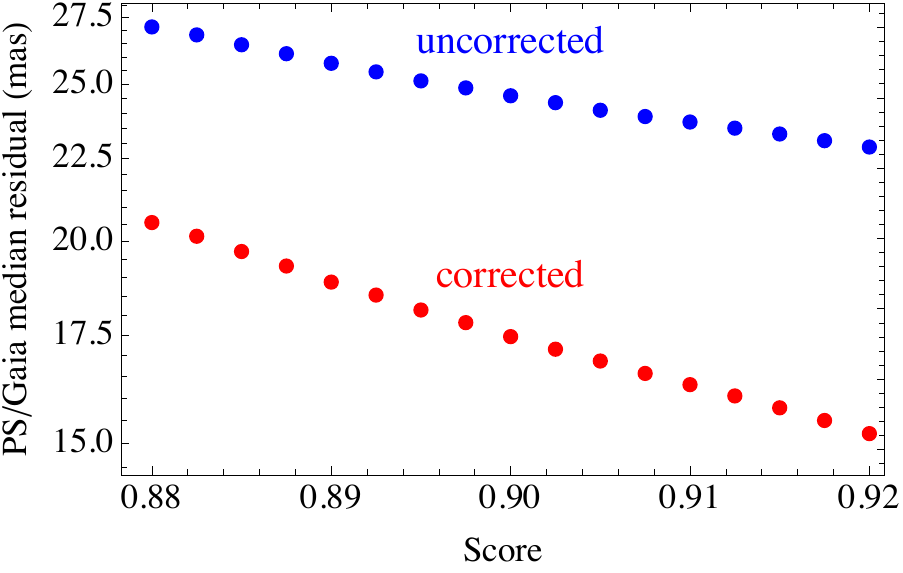}
\caption{Median PS1/Gaia positional residuals in mas as a function of \psscore\ (degree to which the object is point-like)
in a small interval about \psscore=0.9 with and without the corrections described in Section \ref{sec:algorithm}. Only objects with \psscore\ $> 0.9$ are reference objects. }
\label{fig:GlobalScoreDAllTrans}
 \end{figure}

If the residuals are a true measure of error,
they should be continuous across the \psscore\ boundary of 0.9 between reference and non-reference objects. Figure~\ref{fig:GlobalScoreDAllTrans} plots
the residuals in a small \psscore\ interval near this boundary. As seen in the plot, the corrected as well as the uncorrected cases
are in fact smooth functions of \psscore.
(Note that these residuals are larger than those for the reference sample because objects with smaller \psscores\ are resolved or blended in PS1, leading to larger errors in their positions.)

\subsection{Residuals in Low and High Reference Object Density Regions}
\label{sec:dmax}

\begin{figure}
\centering
    \plotone{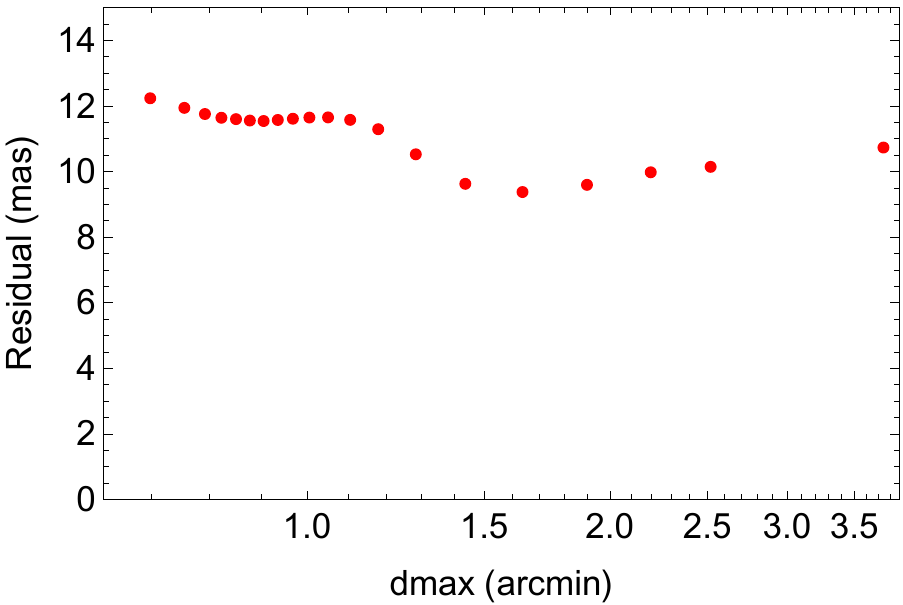}
\caption{Median PS1/Gaia separation distance with corrected astrometry as a function of the maximum distance to the nearest 33 neighboring reference objects (on a logarithmic scale)  for all reference objects in Stripe 2.  Adjacent points in the plot involve an equal number of objects.}
\label{fig:dmaxD}
 \end{figure}

\begin{figure}
\centering
    \plotone{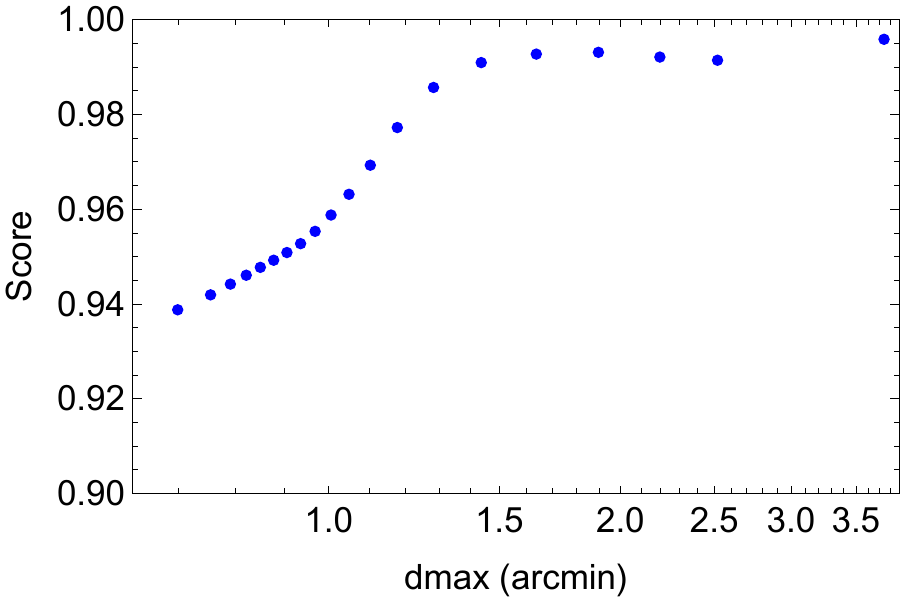}
\caption{Median \psscore\ as a function of the maximum distance to the nearest 33 neighboring reference objects (on a logarithmic scale)  for all reference objects in Stripe 2. Adjacent points in the plot involve an equal number of objects.}
\label{fig:dmaxScore}
 \end{figure}

The nearest 33 neighboring reference objects are typically distributed over a distance
$d_{\rm max} \sim 1$ arcmin from the object being corrected, but there is a considerable range in $d_{\rm max}$ due to reference object density variations (Fig.~\ref{fig:dmax}).  It is possible that
that there could be a degradation of astrometric accuracy in lower density or higher $d_{\rm max}$ cases.
We examine how the
PS1/Gaia residuals vary with $d_{\rm max}$.

To study this correlation, we create a plot by constructing bins in intervals of $d_{\rm max}$  values that contain an equal number of reference objects, as we did for Figure~\ref{fig:dscore}. The results for Stripe 2 ($ -28.3^\circ < \delta < -25.0^\circ$) are plotted in Figure~\ref{fig:dmaxD}.
The separation between adjacent plotted points in Figure~\ref{fig:dmaxD}
reflects the number distribution of reference objects with $d_{\rm max}$. The close
separation near $d_{\rm max} \sim 1$ arcmin  is due to the high number of reference objects near that value, as seen in Figure~\ref{fig:dmax}, while the wide separation near the largest value of $d_{\rm max}$ reflects the rarity of cases that require large $d_{\rm max}$  values
to find 33 neighbors.
The residuals
are quite insensitive to $d_{\rm max}$ and slightly decline with increasing $d_{\rm max}$  near $d_{\rm max} \simeq 1.3$ arcmin.
The median absolute deviation (MAD) values of these residuals follow a similar pattern.
These properties show that the correction algorithm works well in both high and low reference object density regions.

From Figure~\ref{fig:dscore} we know that the astrometric accuracy increases with increasing values of \psscore.
In Figure~\ref{fig:dmaxScore} we see that \psscore\ increases with
increasing $d_{\rm max}$. That is, higher reference object density regions tend
to have fewer point-like object detections than lower density regions.
That is due mainly to increased blending in crowded PS1 fields.
The decline in positional residuals in Figure~\ref{fig:dmaxD} with increasing $d_{\rm max}$
in the corrected case
can be explained in part by the increase in \psscore\ with
increasing $d_{\rm max}$.

\subsection{Comparison With the Hubble Source Catalog}
\label{sec:HSC}

\begin{figure}
\centering
    \plotone{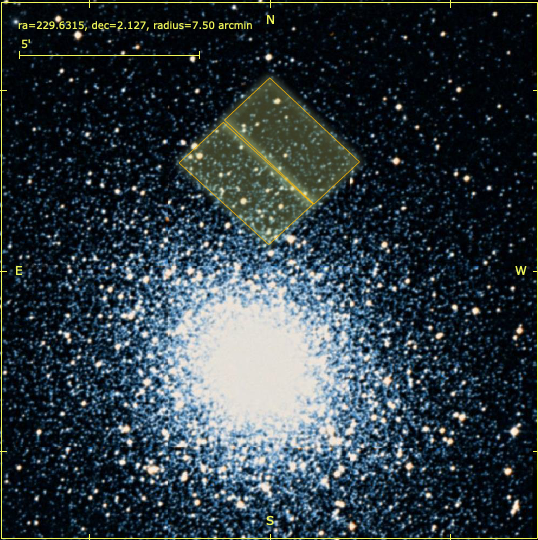}
\caption{Image of globular cluster M5 with footprint of HST\_12517\_13\_ACS\_WFC. }
\label{fig:M5}
 \end{figure}

 \begin{figure}
\centering
    \plotone{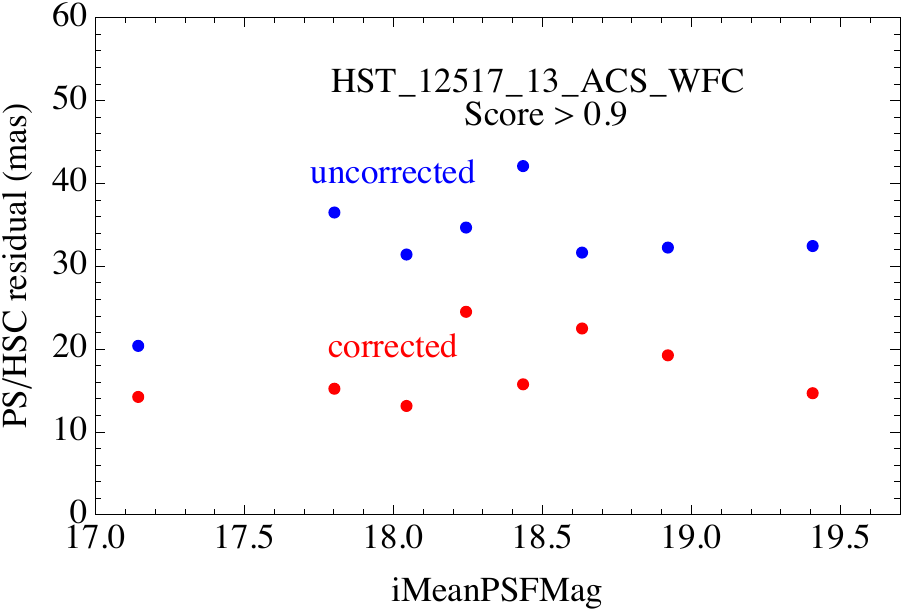}
	 \caption{The median PS1/HSC positional residuals in mas as a function of PS1 \texttt{iMeanPSFMag} for PS1 objects with \psscore\ $> 0.9$. The HSC objects are detected in image HST\_12517\_13\_ACS\_WFC. The number of objects is the same between adjacent points.}
\label{fig:HSCMagDScore}
 \end{figure}

 \begin{figure}
\centering
    \plotone{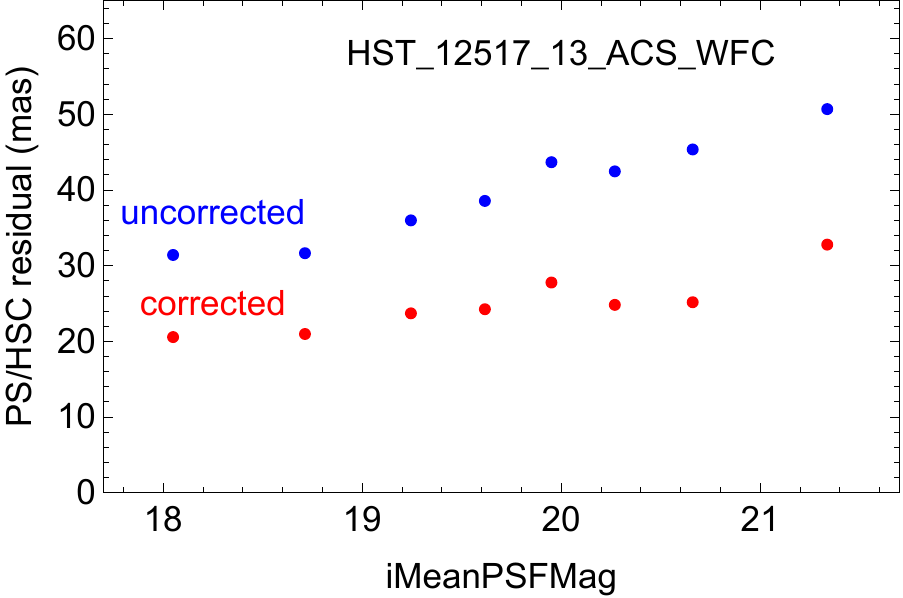}
\caption{Same as Figure~\ref{fig:HSCMagDScore}, but for all \psscore\ values. }
\label{fig:HSCMagD}
 \end{figure}

\begin{figure}
\centering
    \plotone{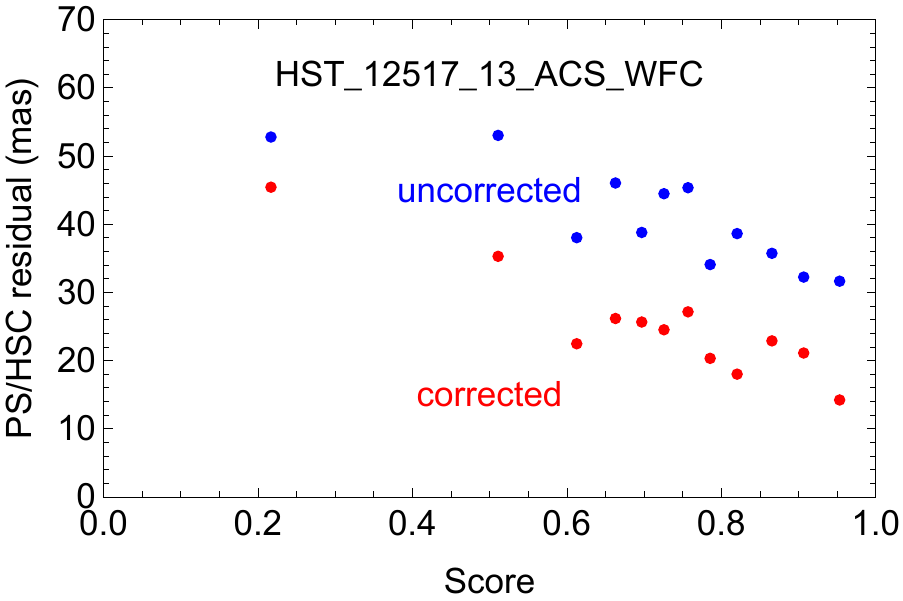}
	\caption{The median PS1/HSC positional residuals in mas as a function of \psscore. Adjacent points in the plot involve an equal number of objects. }
\label{fig:HSCScoreD}
 \end{figure}

The Hubble Source Catalog (HSC) provides high precision astrometry for objects
detected in \textit{Hubble\/} images \citep{Budavari2012, Whitmore2016}.
Although \textit{Hubble\/} images cover a tiny fraction of the sky, they contain much
fainter objects than either PS1 or Gaia. The HSC contains objects fainter than mag 26. We examine the positional residuals
resulting from objects cross-matched between PS1 and the HSC.
We are interested in using HSC objects obtained in an HST image taken close
to the mean PS1 epoch of mid-2012, in order to minimize proper motion shifts.
We also want an HST image with large number of unsaturated HSC detected stars
that cross match with Gaia DR2 to provide accurate astrometry. But the image should
not contain so many stars that crowding in PS1 could be a problem.

For this purpose we used HSC objects detected
in visit level image HST\_12517\_13\_ACS\_WFC that covers a region in the outer parts of globular cluster M5 as seen in Figure~\ref{fig:M5}.  This image consists of a single 100 second exposure in ACS/WFC filter F606W.
The HSC objects in this image
were cross-matched to Gaia DR2 to provide absolute astrometry by using the correction
algorithm described in \cite{Budavari2012}. The image contains 6150 HSC objects,
70 of which were cross-matched with the positions in the Gaia DR2 catalog corrected by proper
motions and parallaxes to the date of the image, June 2012. We note that this astrometric correction to the HSC using Gaia DR2 is not yet publicly available in an HSC release.

The PS1 objects that lie within the image region were cross-matched with
the 6150 HSC objects, resulting in 1089 cross matches
within 0.1 arcsec.
Of these,  111 objects have a \psscore\ greater than 0.9.
Figure~\ref{fig:HSCMagDScore} plots the PS1/HSC residuals based on these 111 objects, both before
and after the PS1 corrections we apply, as a function of PS1 \texttt{iMeanPSFMag}. The median of the residuals before correction is 32 mas and 18 mas after
correction. The values are similar to the median of the residuals before and after correction found in cross matching these same PS1 objects
with Gaia of 29 mas and 15 mas, respectively.
The figure shows that these PS1/HSC cross matches extend to about 19.5 mag, within the Gaia range.

Figure~\ref{fig:HSCMagD} includes all PS1 objects that we consider (i.e., having more than two detections) with all \psscore\ values. In this case, the corrected residuals are larger,
about 40\% larger than in Figure~\ref{fig:HSCMagDScore}. Figure~\ref{fig:HSCScoreD}
shows that the residuals depend on the \psscore\ and grow by more than a factor of 2
from a \psscore\ of 1 to a low \psscore.

 Figure~\ref{fig:HSCMagD}  shows
residuals  to the limiting PS1 magnitude, well beyond the limiting  Gaia
magnitude. The main point is that the reduction
in the PS1 corrected residuals is present beyond the Gaia magnitude limit.
The lower PS1 astrometric accuracy at faint magnitudes is a consequence of the lower signal to noise at fainter
magnitudes, which both increases the positional errors and also makes the identification of point-like objects using
\psscore\ less reliable. Regardless of the challenges near the PS1 detection limit, the corrections we apply to PS1 significantly improve
the astrometric accuracy of such faint objects.

\begin{figure}
	\centering \leavevmode
	\includegraphics[width=0.99\columnwidth]{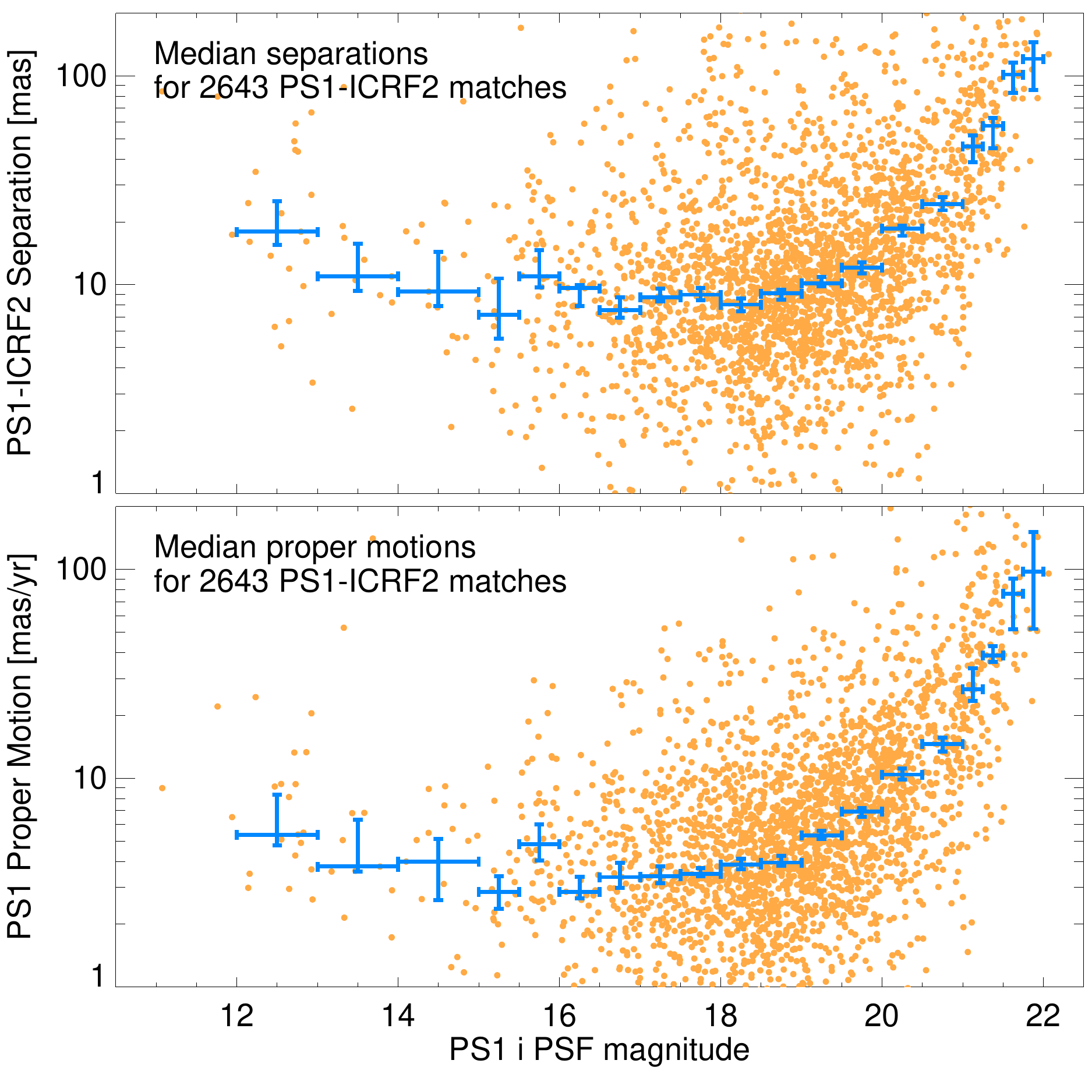}
	\caption{The top panel shows offsets between our improved PS1
	positions and the ICRF2 radio catalog \citep{Ma2009} as a function of
	the PS1 $i$-band PSF magnitude.  The error bars show the medians and
	$1\sigma$ ranges in magnitude bins. The bottom panel shows the PMs for
	the same objects.  Since the ICRF2 objects are extragalactic, they all
	have true proper motions of zero, so this is an estimate of the
	uncertainties of the PS1 proper motions.  The scatter is somewhat
	larger than in the comparisons to Gaia because ICRF2 optical
	counterparts are frequently resolved or blended in the PS1 data.  Note
	that this sample goes deeper than Gaia and can be used for
	positional noise estimates to the PS1 detection limit.
	}
\label{fig:icrf2}
\end{figure}

\subsection{Comparison to the ICRF2 Catalog}
\label{sec:icrf2}

As an independent astrometric test, we cross-matched our PS1 catalog to the
International Celestial Reference Frame catalog \citep[ICRF2;][]{Ma2009}.  The
ICRF2 catalog is the basic astrometric reference catalog defining the radio
coordinate system. It includes 3414 sources spread over the entire sky with
very accurate radio positions having errors typically less than 1 mas.  Since
the ICRF2 objects are all extragalactic sources, they have vanishingly small
proper motions.

A cross-match with PS1 found 2679 objects within 0.5 arcsec of the ICRF2
positions.\footnote{Unmatched objects are mainly below the $-30^\circ$
declination limit, with a few falling into small holes in the PS1 sky
coverage.} Of those, 2643 are detected in the PS1 $i$-band PSF magnitude.  The
positional and proper motion accuracies as function of magnitude are shown in
Fig.~\ref{fig:icrf2}.  The scatter in this comparison is slightly larger than the
PS1/Gaia cross-match (Figs.~\ref{fig:dimag} and \ref{fig:pmimag}) because many ICRF2
optical counterparts are resolved galaxies or are blended with neighboring objects
in the PS1 images.  We have elected not to eliminate objects with \psscore\
less than 0.9 in this comparison because we wish to retain the fainter sources to get
an estimate of the uncertainties below the Gaia magnitude limit.

\cite{Berghea2016} utilized a selected subset of ICRF2 sources
to determine astrometric corrections for the PS1 catalog
through the Global Astrometric Solution (GAS) method. The pre-release version of the PS1 catalog
they used had an astrometric calibration based on 2MASS rather than on Gaia DR1 (which was
not yet available).  Their results removed the large scale
errors in the PS1 DR1 due to its reference catalog. However,
residual errors of about 60 mas remained in RA and declination that
were correlated on small scales of a fraction of a degree.
These small-scale errors are likely related to the errors that are corrected by our algorithm.
Both the positions and proper motions from this paper are much more accurate
than those from \cite{Berghea2016} because the
higher sky density of the Gaia DR2 catalog enables the
small-scale corrections that are required to remove distortions
in the PS1 astrometry.

The comparison to ICRF2 also confirms the accuracy of our PS1 positions and proper motions to magnitudes fainter
than the Gaia DR2 catalog detection limit.  For objects in the magnitude range $17 < i < 19$, the
median error in the PS1 position is 8.7~mas, and the median error in the PM is 3.8~mas/yr.
The errors increase for fainter objects, as expected due to noise in the PS1 images.

\section{Discussion and Summary}
 \label{sec:summary}

 We have carried out a procedure to improve the
astrometric accuracy of about 1.7 billion PS1 objects using Gaia DR2.
In addition, we provide proper motions for these objects that are also corrected with Gaia.
The astrometric correction procedure makes use of Gaia proper motions and parallaxes
to shift the Gaia positions to the epochs of the PS1 objects.
The PS1 DR2 catalog has systematic astrometric errors on a $\sim 1$ arcmin scale that are greatly
reduced by our algorithm.
For a subset of PS1 objects that are point-like (reference objects),
the corrections result in mode (peak) and median  PS1/Gaia positional
residuals of about 4 mas and 9 mas, respectively,
which is about a 33\% reduction for the median  (Fig.~\ref{fig:globald}).
The proper motions for these objects are corrected by a similar procedure to about 2 mas/yr (mode) and 5 mas/yr (median)
compared with Gaia, which represents a 24\% improvement for the median (see Fig.~\ref{fig:globalpm}).
The highest astrometric accuracy occurs for the most point-like objects and
for intermediate magnitudes of about 17 mag (Figs.~\ref{fig:dscore} to \ref{fig:pmimag}).

The positional corrections to a given PS1 object on the  $\sim 1$ arcmin scale involves
taking the median values of the PS1  to Gaia  position shifts of the nearest 33 neighboring cross-matched objects, excluding
the object itself (see Section \ref{sec:algorithm}). In this process we are not force fitting
positions of individual PS1 objects to cross-matched Gaia objects. Instead, we are smoothing spatial structures in the PS1/Gaia positional differences on the $\sim 1$ arcmin scale (Fig.~\ref{fig:dmax}). The correction process converges well with
the number of nearest neighbors. For the 33 nearest neighbors that we apply,
Figure~\ref{fig:conv} in Appendix~\ref{sec:convergence} suggests that
the results are converged to within about 1 mas.
The corrections are applied to all PS1 objects that we consider,
most of which do not cross match with a Gaia object.
Similar corrections are applied to PS1 proper motions,
although the improvements are not as large as for the positions.

The positional residuals in RA and declination are quite different (Fig.~\ref{fig:resstripes}).
The differences are minimized near zero zenith angle (see Fig.~\ref{fig:ramdec}). The difference between the
RA and declination residual distributions, as measured by the MAD of the residuals within stripes,
varies with zenith angle and is  proportional to the air mass
(Fig.~\ref{fig:resmada}). The declination residuals
within stripes  vary with object color while the RA residuals
are nearly independent of color (Fig.~\ref{fig:colormed}).
The declination residual variations with color are much smaller in Stripe 16, which passes through the zenith, than at zenith angle 47 degrees (Stripes 2 and 30). These variations are consistent with the effects
of atmospheric refraction that dispersively bend light rays
by an amount that depends on declination.
In principle, such color dependent effects could be calibrated and removed to
provide a further reduction in residuals.

In the future, we intend to make these astrometry improvements available through MAST.
There are also several areas for future improvements:
\begin{enumerate}

\item While we have corrected the astrometry for 1.7 billion objects which have three or more detection epochs, over 8
billion other objects have two or fewer detection epochs.
That includes many spurious detections, but also includes faint objects that are detected only on the 
stacked images that combine the single-epoch observations.
We plan to apply our astrometric correction method to these objects as well.

\item The systematic color-dependent residuals described in Section \ref{sec:color}
can be used to further improve the astrometric accuracy of the reference stars. The reference stars can in turn
be used to make color-dependent corrections to the other PS1 objects. The inclusion of color-dependent corrections in
our algorithm could result in substantial additional improvements in the astrometry, since the color terms appear large
enough to account for much of the remaining scatter in the astrometric ``sweet spot'' around magnitude 17 (see
Figs.~\ref{fig:colormed} and~\ref{fig:colormad}).

\item Our proper motion and parallax determinations
are limited to smaller shifts over the range of PS1 epochs. The PS1 catalog splits measurements of detections that
move more than 1 arcsec into separate objects, and the 
2 arcsec PS1/Gaia cross matching that we initially apply also limits proper
motions to less than about 0.7 mas/yr.  Given the measured PMs of objects, it is likely that we can identify
some that have multiple entries in the catalog and can re-unite them into single, more accurately measured objects. 

We plan to reapply our corrections to PS1 astrometry as improved versions of the Gaia and PS1 catalogs become available. Our pipeline is well enough automated to make this possible without a lot of effort.

\end{enumerate}

\section*{Acknowledgments}

We thank Mike Fall and the Astrometry Working Group at STScI for 
motivating this work and providing feedback during the preliminary phase.
In particular, we thank Stefano Casertano for examining the level of improvement
that could be possible for PS1 astrometry by using Gaia.
We also thank
Ciprian Berghia, Valeri Makarov, and Juliene Frouard for insightful discussions
about the limitations on the use of Gaia in the current PS1 astrometry.
The Pan-STARRS1 Surveys (PS1) and the PS1 public science archive have been made possible through contributions by the Institute for Astronomy, the University of Hawaii, the Pan-STARRS Project Office, the Max-Planck Society and its participating institutes, the Max Planck Institute for Astronomy, Heidelberg and the Max Planck Institute for Extraterrestrial Physics, Garching, The Johns Hopkins University, Durham University, the University of Edinburgh, the Queen's University Belfast, the Harvard-Smithsonian Center for Astrophysics, the Las Cumbres Observatory Global Telescope Network Incorporated, the National Central University of Taiwan, the Space Telescope Science Institute, the National Aeronautics and Space Administration under Grant No. NNX08AR22G issued through the Planetary Science Division of the NASA Science Mission Directorate, the National Science Foundation Grant No. AST-1238877, the University of Maryland, Eotvos Lorand University (ELTE), the Los Alamos National Laboratory, and the Gordon and Betty Moore Foundation.

\bibliographystyle{aasjournal}
\bibliography{references.bib}

\begin{thebibliography}{}
\expandafter\ifx\csname natexlab\endcsname\relax\def\natexlab#1{#1}\fi
\providecommand{\url}[1]{\href{#1}{#1}}
\providecommand{\dodoi}[1]{doi:~\href{http://doi.org/#1}{\nolinkurl{#1}}}
\providecommand{\doeprint}[1]{\href{http://ascl.net/#1}{\nolinkurl{http://ascl.net/#1}}}
\providecommand{\doarXiv}[1]{\href{https://arxiv.org/abs/#1}{\nolinkurl{https://arxiv.org/abs/#1}}}

\bibitem[{{Berghea} {et~al.}(2016){Berghea}, {Makarov}, {Frouard}, {Hennessy},
  {Dorland}, {Veillette}, {Dudik}, {Magnier}, {Burgett}, {Chambers}, {Denneau},
  {Flewelling}, {Kaiser}, {Tonry}, {Wainscoat}, \& {Sesar}}]{Berghea2016}
{Berghea}, C.~T., {Makarov}, V.~V., {Frouard}, J., {et~al.} 2016, \aj, 152, 53,
  \dodoi{10.3847/0004-6256/152/3/53}

\bibitem[{{Budav{\'a}ri} \& {Lubow}(2012)}]{Budavari2012}
{Budav{\'a}ri}, T., \& {Lubow}, S.~H. 2012, \apj, 761, 188,
  \dodoi{10.1088/0004-637X/761/2/188}

\bibitem[{{Budav{\'a}ri} {et~al.}(2010){Budav{\'a}ri}, {Szalay}, \&
  {Fekete}}]{Budavari2010}
{Budav{\'a}ri}, T., {Szalay}, A.~S., \& {Fekete}, G. 2010, \pasp, 122, 1375,
  \dodoi{10.1086/657302}

\bibitem[{{Chambers} {et~al.}(2016){Chambers}, {Magnier}, {Metcalfe},
  {Flewelling}, {Huber}, {Waters}, {Denneau}, {Draper}, {Farrow}, {Finkbeiner},
  {Holmberg}, {Koppenhoefer}, {Price}, {Rest}, {Saglia}, {Schlafly}, {Smartt},
  {Sweeney}, {Wainscoat}, {Burgett}, {Chastel}, {Grav}, {Heasley}, {Hodapp},
  {Jedicke}, {Kaiser}, {Kudritzki}, {Luppino}, {Lupton}, {Monet}, {Morgan},
  {Onaka}, {Shiao}, {Stubbs}, {Tonry}, {White}, {Ba{\~n}ados}, {Bell},
  {Bender}, {Bernard}, {Boegner}, {Boffi}, {Botticella}, {Calamida},
  {Casertano}, {Chen}, {Chen}, {Cole}, {Deacon}, {Frenk}, {Fitzsimmons},
  {Gezari}, {Gibbs}, {Goessl}, {Goggia}, {Gourgue}, {Goldman}, {Grant},
  {Grebel}, {Hambly}, {Hasinger}, {Heavens}, {Heckman}, {Henderson}, {Henning},
  {Holman}, {Hopp}, {Ip}, {Isani}, {Jackson}, {Keyes}, {Koekemoer}, {Kotak},
  {Le}, {Liska}, {Long}, {Lucey}, {Liu}, {Martin}, {Masci}, {McLean}, {Mindel},
  {Misra}, {Morganson}, {Murphy}, {Obaika}, {Narayan}, {Nieto-Santisteban},
  {Norberg}, {Peacock}, {Pier}, {Postman}, {Primak}, {Rae}, {Rai}, {Riess},
  {Riffeser}, {Rix}, {R{\"o}ser}, {Russel}, {Rutz}, {Schilbach}, {Schultz},
  {Scolnic}, {Strolger}, {Szalay}, {Seitz}, {Small}, {Smith}, {Soderblom},
  {Taylor}, {Thomson}, {Taylor}, {Thakar}, {Thiel}, {Thilker}, {Unger},
  {Urata}, {Valenti}, {Wagner}, {Walder}, {Walter}, {Watters}, {Werner},
  {Wood-Vasey}, \& {Wyse}}]{Chambers2016}
{Chambers}, K.~C., {Magnier}, E.~A., {Metcalfe}, N., {et~al.} 2016, arXiv
  e-prints, arXiv:1612.05560.
\newblock \doarXiv{1612.05560}

\bibitem[{{Cudworth} \& {Rees}(1990)}]{Cu1990}
{Cudworth}, K.~M., \& {Rees}, R. 1990, \aj, 99, 1491, \dodoi{10.1086/115434}

\bibitem[{{Flewelling} {et~al.}(2016){Flewelling}, {Magnier}, {Chambers},
  {Heasley}, {Holmberg}, {Huber}, {Sweeney}, {Waters}, {Calamida}, {Casertano},
  {Chen}, {Farrow}, {Hasinger}, {Henderson}, {Long}, {Metcalfe}, {Narayan},
  {Nieto-Santisteban}, {Norberg}, {Rest}, {Saglia}, {Szalay}, {Thakar},
  {Tonry}, {Valenti}, {Werner}, {White}, {Denneau}, {Draper}, {Hodapp},
  {Jedicke}, {Kaiser}, {Kudritzki}, {Price}, {Wainscoat}, {Builders},
  {Chastel}, {McLean}, {Postman}, \& {Shiao}}]{Flewelling2016}
{Flewelling}, H.~A., {Magnier}, E.~A., {Chambers}, K.~C., {et~al.} 2016, arXiv
  e-prints, arXiv:1612.05243.
\newblock \doarXiv{1612.05243}

\bibitem[{{Heasley}(2008)}]{Heasley2008}
{Heasley}, J.~N. 2008, in American Institute of Physics Conference Series, Vol.
  1082, American Institute of Physics Conference Series, ed. C.~A.~L.
  {Bailer-Jones}, 352--358, \dodoi{10.1063/1.3059075}

\bibitem[{{Lindegren} {et~al.}(2016){Lindegren}, {Lammers}, {Bastian},
  {Hern{\'a}ndez}, {Klioner}, {Hobbs}, {Bombrun}, {Michalik}, {Ramos-Lerate},
  {Butkevich}, {Comoretto}, {Joliet}, {Holl}, {Hutton}, {Parsons},
  {Steidelm{\"u}ller}, {Abbas}, {Altmann}, {Andrei}, {Anton}, {Bach},
  {Barache}, {Becciani}, {Berthier}, {Bianchi}, {Biermann}, {Bouquillon},
  {Bourda}, {Br{\"u}semeister}, {Bucciarelli}, {Busonero}, {Carlucci},
  {Casta{\~n}eda}, {Charlot}, {Clotet}, {Crosta}, {Davidson}, {de Felice},
  {Drimmel}, {Fabricius}, {Fienga}, {Figueras}, {Fraile}, {Gai}, {Garralda},
  {Geyer}, {Gonz{\'a}lez-Vidal}, {Guerra}, {Hambly}, {Hauser}, {Jordan},
  {Lattanzi}, {Lenhardt}, {Liao}, {L{\"o}ffler}, {McMillan}, {Mignard}, {Mora},
  {Morbidelli}, {Portell}, {Riva}, {Sarasso}, {Serraller}, {Siddiqui}, {Smart},
  {Spagna}, {Stampa}, {Steele}, {Taris}, {Torra}, {van Reeven}, {Vecchiato},
  {Zschocke}, {de Bruijne}, {Gracia}, {Raison}, {Lister}, {Marchant},
  {Messineo}, {Soffel}, {Osorio}, {de Torres}, \& {O'Mullane}}]{Lindegren2016}
{Lindegren}, L., {Lammers}, U., {Bastian}, U., {et~al.} 2016, \aap, 595, A4,
  \dodoi{10.1051/0004-6361/201628714}

\bibitem[{{Lindegren} {et~al.}(2018){Lindegren}, {Hern{\'a}ndez}, {Bombrun},
  {Klioner}, {Bastian}, {Ramos-Lerate}, {de Torres}, {Steidelm{\"u}ller},
  {Stephenson}, {Hobbs}, {Lammers}, {Biermann}, {Geyer}, {Hilger}, {Michalik},
  {Stampa}, {McMillan}, {Casta{\~n}eda}, {Clotet}, {Comoretto}, {Davidson},
  {Fabricius}, {Gracia}, {Hambly}, {Hutton}, {Mora}, {Portell}, {van Leeuwen},
  {Abbas}, {Abreu}, {Altmann}, {Andrei}, {Anglada}, {Balaguer-N{\'u}{\~n}ez},
  {Barache}, {Becciani}, {Bertone}, {Bianchi}, {Bouquillon}, {Bourda},
  {Br{\"u}semeister}, {Bucciarelli}, {Busonero}, {Buzzi}, {Cancelliere},
  {Carlucci}, {Charlot}, {Cheek}, {Crosta}, {Crowley}, {de Bruijne}, {de
  Felice}, {Drimmel}, {Esquej}, {Fienga}, {Fraile}, {Gai}, {Garralda},
  {Gonz{\'a}lez-Vidal}, {Guerra}, {Hauser}, {Hofmann}, {Holl}, {Jordan},
  {Lattanzi}, {Lenhardt}, {Liao}, {Licata}, {Lister}, {L{\"o}ffler},
  {Marchant}, {Martin-Fleitas}, {Messineo}, {Mignard}, {Morbidelli}, {Poggio},
  {Riva}, {Rowell}, {Salguero}, {Sarasso}, {Sciacca}, {Siddiqui}, {Smart},
  {Spagna}, {Steele}, {Taris}, {Torra}, {van Elteren}, {van Reeven}, \&
  {Vecchiato}}]{Lindegren2018}
{Lindegren}, L., {Hern{\'a}ndez}, J., {Bombrun}, A., {et~al.} 2018, \aap, 616,
  A2, \dodoi{10.1051/0004-6361/201832727}

\bibitem[{{Ma} {et~al.}(2009){Ma}, {Arias}, {Bianco}, {Boboltz}, {Bolotin},
  {Charlot}, {Engelhardt}, {Fey}, {Gaume}, {Gontier}, {Heinkelmann}, {Jacobs},
  {Kurdubov}, {Lambert}, {Malkin}, {Nothnagel}, {Petrov}, {Skurikhina},
  {Sokolova}, {Souchay}, {Sovers}, {Tesmer}, {Titov}, {Wang}, {Zharov},
  {Barache}, {Boeckmann}, {Collioud}, {Gipson}, {Gordon}, {Lytvyn},
  {MacMillan}, \& {Ojha}}]{Ma2009}
{Ma}, C., {Arias}, E.~F., {Bianco}, G., {et~al.} 2009, IERS Technical Note, 35,
  1

\bibitem[{{Magnier} {et~al.}(2016){Magnier}, {Schlafly}, {Finkbeiner}, {Tonry},
  {Goldman}, {R{\"o}ser}, {Schilbach}, {Chambers}, {Flewelling}, {Huber},
  {Price}, {Sweeney}, {Waters}, {Denneau}, {Draper}, {Hodapp}, {Jedicke},
  {Kudritzki}, {Metcalfe}, {Stubbs}, \& {Wainscoast}}]{Magnier2016}
{Magnier}, E.~A., {Schlafly}, E.~F., {Finkbeiner}, D.~P., {et~al.} 2016, arXiv
  e-prints, arXiv:1612.05242.
\newblock \doarXiv{1612.05242}

\bibitem[{{Makarov} {et~al.}(2017){Makarov}, {Berghea}, \&
  {Frouard}}]{Makarov2017}
{Makarov}, V., {Berghea}, C.~T., \& {Frouard}, J. 2017, AADD USNO

\bibitem[{{Schubert} \& {Walterscheid}(2000)}]{Schubert2000}
{Schubert}, G., \& {Walterscheid}, R.~L. 2000, {Earth}, ed. A.~N. {Cox}, 239

\bibitem[{{Skrutskie} {et~al.}(2006){Skrutskie}, {Cutri}, {Stiening},
  {Weinberg}, {Schneider}, {Carpenter}, {Beichman}, {Capps}, {Chester},
  {Elias}, {Huchra}, {Liebert}, {Lonsdale}, {Monet}, {Price}, {Seitzer},
  {Jarrett}, {Kirkpatrick}, {Gizis}, {Howard}, {Evans}, {Fowler}, {Fullmer},
  {Hurt}, {Light}, {Kopan}, {Marsh}, {McCallon}, {Tam}, {Van Dyk}, \&
  {Wheelock}}]{Skrutskie2006}
{Skrutskie}, M.~F., {Cutri}, R.~M., {Stiening}, R., {et~al.} 2006, \aj, 131,
  1163, \dodoi{10.1086/498708}

\bibitem[{{Tachibana} \& {Miller}(2018)}]{Tachibana2018}
{Tachibana}, Y., \& {Miller}, A.~A. 2018, \pasp, 130, 128001,
  \dodoi{10.1088/1538-3873/aae3d9}

\bibitem[{{Thakar} {et~al.}(2003){Thakar}, {Szalay}, {Vandenberg}, {Gray}, \&
  {Stoughton}}]{Thakar2003}
{Thakar}, A.~R., {Szalay}, A.~S., {Vandenberg}, J.~V., {Gray}, J., \&
  {Stoughton}, A.~S. 2003, Astronomical Society of the Pacific Conference
  Series, Vol. 295, {Data Organization in the SDSS Data Release 1}, ed. H.~E.
  {Payne}, R.~I. {Jedrzejewski}, \& R.~N. {Hook}, 217

\bibitem[{{Tonry} {et~al.}(2012){Tonry}, {Stubbs}, {Lykke}, {Doherty},
  {Shivvers}, {Burgett}, {Chambers}, {Hodapp}, {Kaiser}, {Kudritzki},
  {Magnier}, {Morgan}, {Price}, \& {Wainscoat}}]{Tonry2012}
{Tonry}, J.~L., {Stubbs}, C.~W., {Lykke}, K.~R., {et~al.} 2012, \apj, 750, 99,
  \dodoi{10.1088/0004-637X/750/2/99}

\bibitem[{{Wallace}(2018)}]{Wa2018}
{Wallace}, J.~J. 2018, Research Notes of the American Astronomical Society, 2,
  213, \dodoi{10.3847/2515-5172/aaf1a2}

\bibitem[{{Whitmore} {et~al.}(2016){Whitmore}, {Allam}, {Budav{\'a}ri},
  {Casertano}, {Downes}, {Donaldson}, {Fall}, {Lubow}, {Quick}, {Strolger},
  {Wallace}, \& {White}}]{Whitmore2016}
{Whitmore}, B.~C., {Allam}, S.~S., {Budav{\'a}ri}, T., {et~al.} 2016, \aj, 151,
  134, \dodoi{10.3847/0004-6256/151/6/134}

\end{thebibliography}

\appendix

\section{Convergence of Position Shifts}
\label{sec:convergence}

\begin{figure}
\centering
\includegraphics[width=0.95\columnwidth]{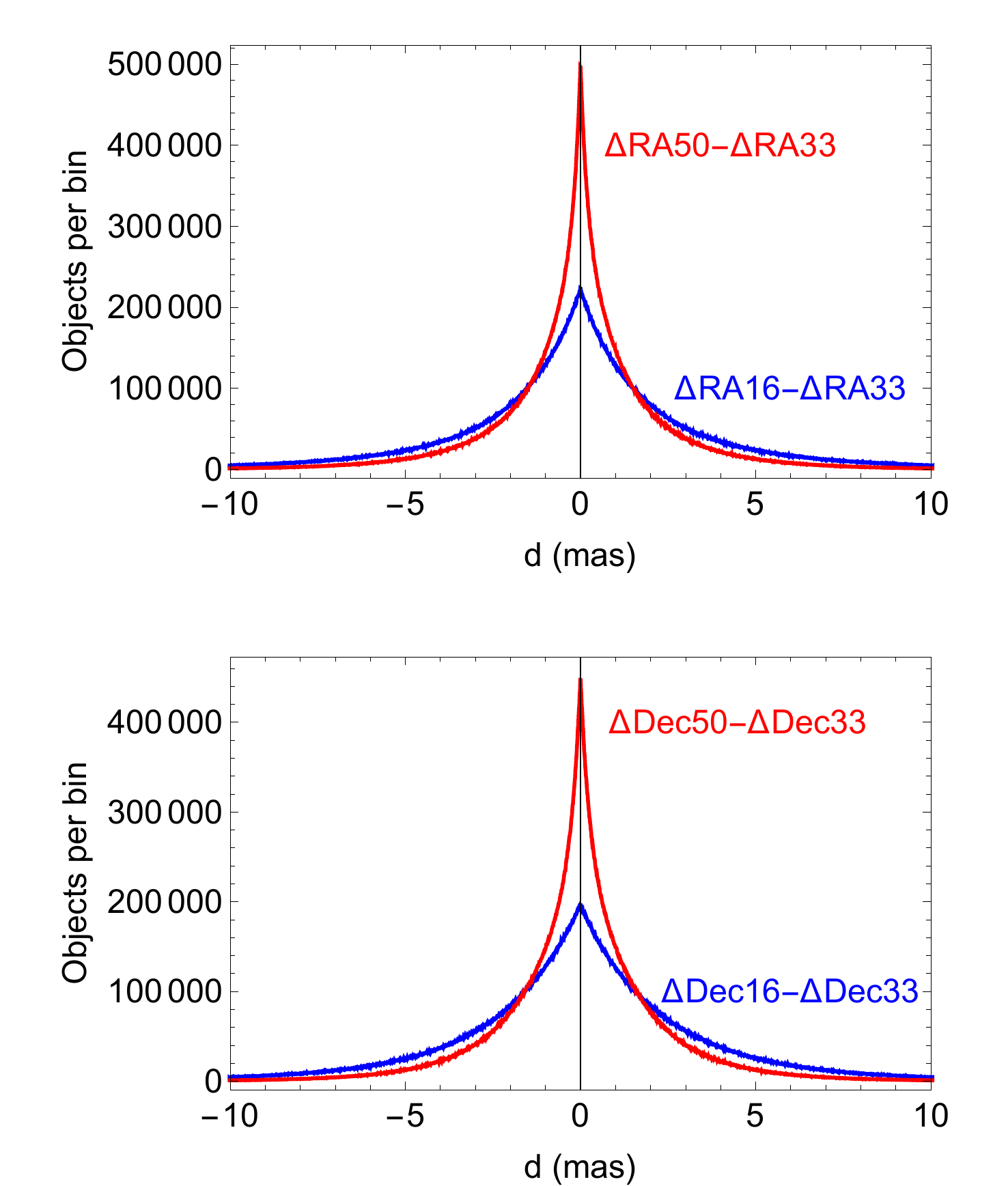}
\caption{Object number distribution in Stripe 2 for the difference in shifts determined by using a different number of nearest neighbors. For example, the upper curve in the upper plot shows the object number distribution for the difference in RA shifts between a calculation using 50 nearest neighbors and a calculation using 33 nearest neighbors. The bin size is 0.01 mas.}
\label{fig:conv}
 \end{figure}

We check that the corrected results are accurately determined using 33 nearest neighbors. To do this, we compare the corrected shifts
in Stripe 2 ($-28.25^\circ < \delta < -25^\circ$) determined by 33 nearest neighbors with shifts resulting from using 16 and 50 nearest neighbors. For each corrected PS1
object, we compute the difference in shifts $\Delta$RA16 - $\Delta$RA33 that is the difference in the RA shift determinations using 16 and 33 neighbors.
Similar calculations are done for the case of 50 nearest neighbors and for the declination. The distributions of these shift differences are plotted in Figure~\ref{fig:conv}.

\begin{figure}
\centering
\includegraphics[width=0.95\columnwidth]{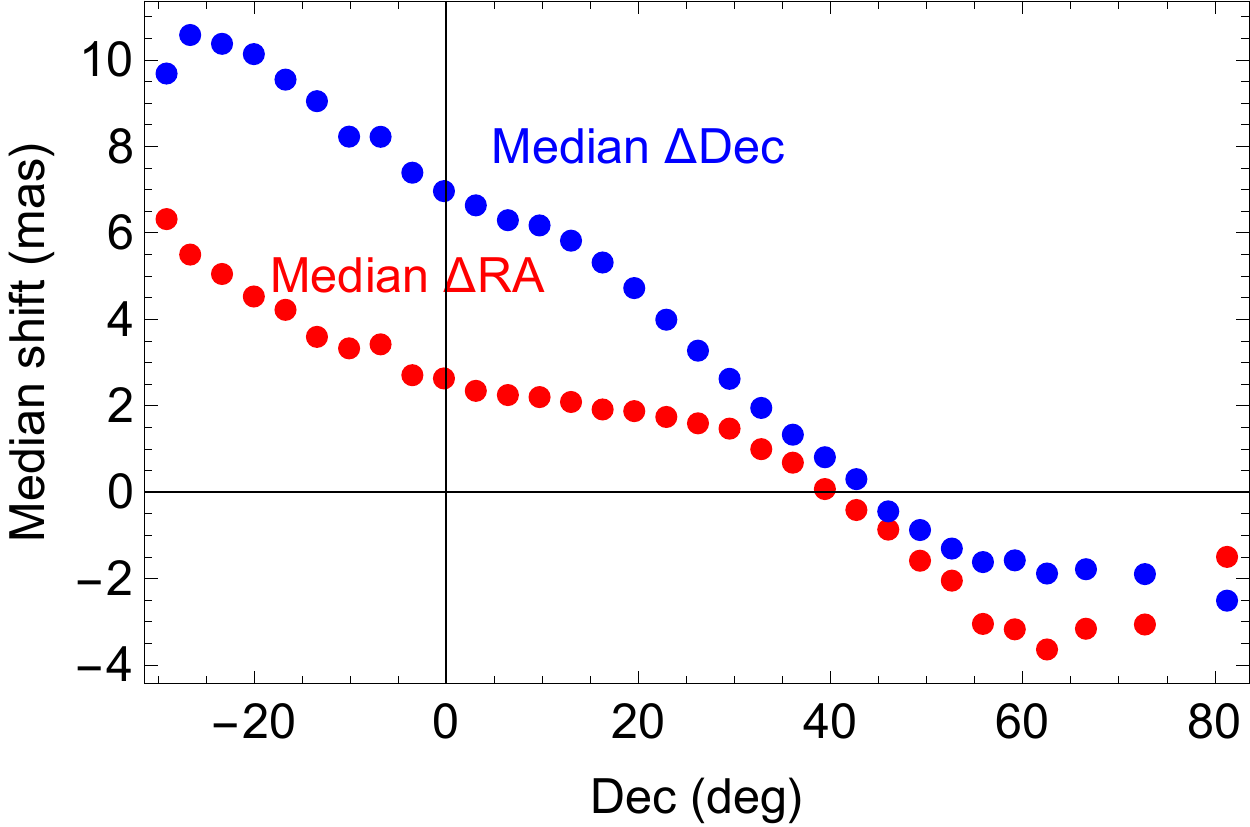}
\caption{Median position shift as a function of declination with one point per stripe.}
\label{fig:medshift}
 \end{figure}

As expected, the distributions are narrower for the cases involving 50 neighbors than for 16 neighbors. The median absolute deviations (MAD) of the
RA distributions are 1.6 mas and 0.9 mas for $\Delta$RA16 - $\Delta$RA33 and $\Delta$RA50 - $\Delta$RA33, respectively. The MADs of the Dec
distributions are 1.8 mas and 1.0 mas for $\Delta$Dec16 - $\Delta$Dec33 and $\Delta$Dec50 - $\Delta$Dec33, respectively. These differences
are much smaller than the typical position shifts, as  seen in Figure~\ref{fig:medshift}. The MADs of $\Delta$RA50 - $\Delta$RA33 and $\Delta$Dec50 - $\Delta$Dec33, about 1 mas, give
a measure of the uncertainty of shifts using the 33 nearest neighbors.

\clearpage

\section{Declination Ranges of Stripes}
\label{sec:Stripes}

PS1 data is partitioned into 32 stripes based on source declination.
Table \ref{tab:stripes} lists in columns
the stripe number, the minimum declination, the maximum
declination, and the average declination of the sources for each stripe.
The declination values are in degrees.

\begin{deluxetable}{crrr}[b]
\tablecaption{PS1 Declination Stripes\label{tab:stripes}}
\tablehead{
	\colhead{Stripe} & \colhead{Min $\delta$} & \colhead{Max $\delta$} & \colhead{Avg $\delta$} \\
	\colhead{~} & \colhead{$^\circ$} & \colhead{$^\circ$} & \colhead{$^\circ$}
}
\tablewidth{0pt}
\startdata
1& -36.58& -28.25& -32.42\\
2& -28.25& -24.99& -26.62\\
3& -25.00& -21.67& -23.33\\
4& -21.67& -18.33& -20.00\\
5& -18.33& -15.08& -16.71\\
6& -15.08& -11.75& -13.42\\
7& -11.75& -8.50& -10.12\\
8& -8.50& -5.17& -6.83\\
9& -5.17& -1.83& -3.50\\
10& -1.84& 1.42& -0.21\\
11& 1.42& 4.75& 3.08\\
12& 4.75& 8.08& 6.42\\
13& 8.08& 11.34& 9.71\\
14& 11.33& 14.67& 13.00\\
15& 14.66& 17.92& 16.29\\
16& 17.92& 21.25& 19.58\\
17& 21.23& 24.58& 22.91\\
18& 24.58& 27.84& 26.21\\
19& 27.83& 31.17& 29.50\\
20& 31.16& 34.42& 32.79\\
21& 34.42& 37.75& 36.08\\
22& 37.75& 41.09& 39.42\\
23& 41.08& 44.33& 42.71\\
24& 44.33& 47.67& 46.00\\
25& 47.67& 51.00& 49.33\\
26& 51.00& 54.25& 52.63\\
27& 54.25& 57.58& 55.92\\
28& 57.58& 60.83& 59.21\\
29& 60.83& 64.33& 62.58\\
30& 64.33& 69.25& 66.79\\
31& 69.25& 77.42& 73.33\\
32& 77.41& 89.99& 83.70\\
\enddata
\end{deluxetable}

\end{document}